\documentclass[a4paper,11pt]{article}
\usepackage{jcappub}

\usepackage{graphicx}
\usepackage{xcolor}
\usepackage{amssymb}
\usepackage{bm}
\usepackage{braket}
\usepackage{aas_macros}

\begin{document}
\title{Parity-violating scalar trispectrum from helical primordial magnetic fields}

\author[a]{Kaito Yura,}
\author[b,c]{Shohei Saga,}
\author[d]{Maresuke Shiraishi,}
\author[a,c,e]{and Shuichiro Yokoyama}

\affiliation[a]{Department of Physics, Nagoya University, 
Furo-cho Chikusa-ku,
Nagoya 464-8602, Japan}
\affiliation[b]{Institute for Advanced Research, Nagoya University,
Furo-cho Chikusa-ku, 
Nagoya 464-8601, Japan}
\affiliation[c]{Kobayashi Maskawa Institute, Nagoya University, Furo-cho Chikusa-ku, Nagoya 464-8602, Japan}
\affiliation[d]{School of General and Management Studies, Suwa University of Science, Chino, Nagano
391-0292, Japan}
\affiliation[e]{Kavli Institute for the Physics and Mathematics of the Universe (WPI), UTIAS, The University of Tokyo, Kashiwa, Chiba 277-8583, Japan}
\emailAdd{yura.kaito.p8@s.mail.nagoya-u.ac.jp}
\emailAdd{shohei.saga@yukawa.kyoto-u.ac.jp}
\emailAdd{shiraishi\_maresuke@rs.sus.ac.jp}
\emailAdd{shu@kmi.nagoya-u.ac.jp}

\date{\today}

\abstract{
Some recent observations of the cosmic microwave background (CMB) anisotropies and the large-scale structure of the Universe imply cosmic parity violation. Among possible parity-violating sources, helical primordial magnetic fields (PMFs) are of particular interest, as they inherently violate parity symmetry and can explain the observed magnetic fields, especially in void regions. PMFs, if generated in the early universe, can source curvature perturbations, which evolve into the present density fluctuations observed in CMB and galaxy surveys. Motivated by this, we study the imprint of helical PMFs on the trispectrum of the sourced primordial curvature perturbations, which is a leading-order scalar statistics sensitive to parity-violating signals.
We derive full expressions for the trispectrum of the primordial curvature perturbations sourced by both the helical and non-helical PMFs and reduce them to computationally-feasible ones using a proper approximation.
From numerical works, we confirm that parity-odd signals are efficiently enhanced and surpass parity-even ones in specific momentum and parameter spaces.
Parity-violating signatures found in this paper are partially testable with observational implications reported so far.
Assuming nearly scale-invariant PMF power spectra and the PMF strength of $B_{r}=4.7 \, {\rm nG}$, we obtain a rough upper bound on the helical-to-non-helical power ratio as $r_H\lesssim 4\times 10^{-4}$.
Our findings highlight the primordial trispectrum as a promising probe of helical PMFs and provide a theoretical basis for future precise observations of higher-order statistics in the CMB anisotropies and the galaxy clustering.}

\maketitle
\flushbottom

\section{Introduction}\label{Sec: Introduction}

The detection of a parity-violating signal on cosmological scales is widely regarded as a smoking gun for new physics beyond the standard cosmological model, which respects parity symmetry. Various cosmological probes have been exploited to hunt parity violation. The cosmic microwave background (CMB) temperature anisotropies and polarizations have provided a robust probe of parity violation. As statistics sensitive to the parity-violating physics in the CMB observations, the two-point cross-correlations between the temperature or $E$-mode polarization, and $B$-mode polarization have been discussed~\cite{Lue:1998mq}. Recent studies have reported a possible detection of $EB$ correlations, with a significance level exceeding $3\sigma$ based on the combined analyses of \textit{Planck} and WMAP data \cite{PhysRevLett.125.221301, PhysRevLett.128.091302, 2022A&A...662A..10E, PhysRevD.106.063503, 2023A&A...679A.144E}. To explain this result, in the context of the so-called cosmic birefringence~\cite{Carroll:1998zi, Lue:1998mq}, the rotation of the polarization plane in propagation, many theoretical models based on parity-violating physics in the late-time universe have been investigated (see, e.g., Ref.~\cite{Fujita:2020ecn}). Beyond two-point statistics, as a probe of the parity-violating phenomenon in the early universe, e.g., the production of the primordial chiral gravitational waves, the bispectrum of the CMB temperature and $E/B$-mode polarization anisotropies has also been studied (see Refs.~\cite{Philcox:2023xxk,Philcox:2024wqx} and references therein).

In parallel, parity-violating signals have also been investigated in the large-scale structure of the Universe. 
For the galaxy clustering in three-dimensional space, which should be a good tracer of the primordial density fluctuations, two- and three-point statistics are insensitive to the parity violation because they are almost invariant under the parity transformation in observed nearly isotropic Universe. This is due to an equivalence between the parity and rotational transformations of scalar two- and three-point statistics. Therefore, four-point statistics, such as the trispectrum in Fourier space, are the lowest-order ones sensitive to parity-violating signatures~\cite{2016PhRvD..94h3503S}. 
Originally, the trispectrum as a probe of the parity violation in the scalar sector has been actively studied in CMB temperature and $E$-mode polarization observations~\cite{2016PhRvD..94h3503S, Philcox:2023ffy, Philcox:2023ypl, Philcox:2025bvj,Philcox:2025lrr,Philcox:2025wts}.
Remarkably, recent analyses of the four-point correlation function of galaxy number density fields have reported the evidence for parity-violating signatures at more than $3\sigma$ significance using the SDSS Baryon Oscillation Spectroscopic Survey data~\cite{2021arXiv210801670P, PhysRevD.106.063501, PhysRevLett.130.201002, 2023MNRAS.522.5701H, PhysRevD.107.023523, Creque-Sarbinowski_2023}. While potential systematic effects, e.g., the underestimation of covariance in mock catalogs, have been pointed out~\cite{2025RSPTA.38340034P, 2024JCAP...08..044K}, it is expected that future observations can give a hint on the new physics associated with the parity violation. In fact, based on these results, the study on the four-point correlations of primordial curvature perturbations generated in the parity-violating models in the early universe has received increasing attention~\cite{Cabass:2022oap, Creque-Sarbinowski:2023wmb, 2023JCAP...05..018N, 2024JHEP...05..196S, 2024JCAP...05..127F, 2024arXiv241011801M}.

Primordial magnetic fields (PMFs), which are considered to be generated before the formation of stars or galaxies, offer a particularly intriguing mechanism for sourcing parity-violating signals in cosmological observables. The presence of magnetic fields with coherence scales of several Mpc in void regions, inferred from the propagation of TeV photons from distant blazars \cite{2010Sci...328...73N, 2011MNRAS.414.3566T, 2012ApJ...747L..14V, 2013ApJ...771L..42T, Yang_2015, 2017ApJ...847...39V}, suggests the existence of a primordial magnetic seed field. In addition, the recent study on the gamma-ray arrival directions observed by Fermi-LAT indicates the non-vanishing parity-odd signal of magnetic fields, so-called helical magnetic fields, in intergalactic regions~\cite{Tashiro:2014gfa,Chen:2014qva}.
These helical fields can originate during inflation and have been shown to explain a plausible origin of observed magnetic fields~\cite{2009JCAP...11..001C, 2014JCAP...10..056C, 2018CQGra..35l4003C, 2018PhRvD..97h3503S, 2022PDU....3601025B, 2022Univ....8...26C, 2023PDU....4001212C}. Their observational signatures have been widely argued and partially tested with observational data, while all previous works are at two- or three-point statistics level (see e.g., Refs.~\cite{Pogosian:2001np, Caprini:2003vc, Kahniashvili:2005xe, 
2012PhRvD..85h3004K, 2012JCAP...06..015S, Ballardini:2014jta, Planck:2015zrl}).

Motivated by these recent investigations, for the first time, we study the parity-odd component of the scalar trispectrum sourced by PMFs, with a particular focus on their helical contributions. Building on previous work that examined helical PMFs in the CMB bispectrum \cite{2012JCAP...06..015S}, we compute the curvature perturbation trispectrum induced by the scalar anisotropic stress of PMFs, commonly referred to as the passive scalar mode. Our approach incorporates both non-helical and helical components of the PMF power spectrum, assuming nearly scale-invariant shape, and exploits the pole approximation introduced in \cite{2012JCAP...06..015S} to make the analytical calculation tractable. 
Through numerical analyses, we find that parity-breaking signatures are efficiently enhanced in specific momentum and parameter spaces. They are partially tested with recent observational implications.

This paper is organized as follows. In Sec.~\ref{Sec: Passive scalar mode and power spectra of primordial magnetic fields}, we introduce the curvature perturbations induced by PMFs, known as the passive scalar mode, and define the power spectrum and the anisotropic stress of the PMFs. In Sec.~\ref{Sec: The analytical results of the passive scalar mode trispectrum}, we explain how to investigate the parity violation by the trispectrum of the curvature perturbation and derive some formulae of the trispectrum of the passive scalar mode by using the pole approximation. In Sec.~\ref{Sec: Numerical results of the passive scalar mode trispectrum}, we numerically validate the pole approximation, and investigate the behavior of the trispectrum from the passive mode in the angular parameter space, assuming the equilateral configuration. In Sec.~\ref{Sec: observational constraint}, we apply the {\it Planck} constraint on the parity-odd amplitudes of the CMB trispectrum to the template of the pole approximation and obtain the upper bound on the ratio of the helical amplitude to non-helical amplitude of PMF power spectrum.
We conclude in Sec.~\ref{Sec: conclusion}.
In Appendix~\ref{sec: app pole full expresions}, we present the full expressions of the trispectrum in the pole approximation. 
In Appendix~\ref{sec: app varying nB}, we study the validity of the approximation we employed in the main text, by varying the spectral indices of the power spectrum of PMFs.

\section{Passive scalar mode and power spectra of primordial magnetic fields}\label{Sec: Passive scalar mode and power spectra of primordial magnetic fields}

In this section, we briefly introduce the curvature perturbation induced by PMFs, which is called the passive scalar mode~\cite{2010PhRvD..81d3517S}. 
In the early universe, when the PMFs are generated by some mechanism, the anisotropic stress of PMFs contributes to the energy-momentum tensor, which acts as a source of the curvature perturbation on super-horizon scales. After the neutrino decoupling, as the non-vanishing anisotropic stress of neutrinos compensates that of the PMFs, the net anisotropic stress in the energy-momentum tensor vanishes. As a result, the curvature perturbation can grow on super-horizon scales between the generation epoch of PMFs and the neutrino decoupling epoch. 
The expression of the resultant curvature perturbation in Fourier space $\zeta_B(\bm{k})$ is approximately given by~\cite{2010PhRvD..81d3517S}
\begin{align} 
\zeta_B(\bm{k})& \simeq
-\frac{3}{2}R_{\gamma}\ln{(\tau_{\nu}/\tau_B)}\Pi_B(\bm{k}), \notag \\
& \equiv \mathcal{T}\,  \Pi_B(\bm{k}),
\label{eq: passive mode curvature perturbation}
\end{align}
where the parameters $\tau_B$, $\tau_{\nu}$, $R_{\gamma}$ are the conformal time at the production of PMFs, the conformal time at the neutrino decoupling, and the ratio between the energy densities of photons and all relativistic particles for $\tau <\tau_{\nu}$, respectively. Throughout this paper, 
we take $\tau_{\nu}/\tau_B=10^{17}$~\cite{2012JCAP...06..015S}.
Here, the quantity $\Pi_{B}(\bm{k})$ is the scalar part of the anisotropic stress of PMFs, which is defined by using the scalar projection tensor $O^{(0)}_{ij}(\hat{\bm{k}}) \equiv \delta_{ij}/3 - \hat{k}_{i}\hat{k}_{j}$ as
\begin{align}
\Pi_{B}(\bm{k}) &= O^{(0)}_{ij}(\hat{\bm{k}}) \Pi_{B\, ij}(\bm{k}) , \label{eq: scalar PiB}
\end{align}
where $\hat{\bm{k}}$ denotes the unit vector of $\bm{k}$, and we adopt the Einstein summation convention for the subscripts $i,j$ which run from 1 to 3.
The function $\Pi_{B\, ij}(\bm{k})$ is the anisotropic stress tensor of PMFs given by
\begin{align}
\Pi_{B\, ij}(\bm{k}) = -\frac{1}{4\pi \rho_{\gamma,0}}\int\frac{{\rm d}^3\bm{k}'}{(2\pi)^3}B_i(\bm{k}')B_j(\bm{k}-\bm{k}'), \label{eq: pi B}
\end{align}
with $\rho_{\gamma,0}$ being the present value of the photon energy density. We note that the quantity $\bm{B}(\bm{k}) = \bm{B}(\tau,\bm{k})a^{2}(\tau)$, with $a$ being a scale factor, is the comoving magnetic field in Fourier space, i.e., the magnetic fields at the present epoch assuming the adiabatic decay $\propto a^{-2}$ due to the expansion of the universe.

Assuming the Gaussianity of PMFs, the statistical properties of the PMFs are fully captured by using the power spectrum:~\cite{Caprini:2003vc,2012JCAP...06..015S}
\begin{align}
\langle B_i(\bm{k})B_j(\bm{k}')\rangle&=\frac{(2\pi)^3}{2}\delta_{\mathrm{D}}(\bm{k}+\bm{k}')[P_B(k)P_{ij}(\hat{\bm{k}})+i\eta_{ija}\hat{k}^aP_{H}(k)]
\label{power spectrum of PMFs}
\\
&\equiv
\frac{(2\pi)^3}{2}\delta_{\mathrm{D}}(\bm{k}+\bm{k}')\mathcal{P}_{ij}(\bm{k}),
\end{align}
where $\delta_{\mathrm{D}}$ is the Dirac delta function, $\eta_{ija}$ is the Levi-Civita tensor, and $P_{ij}(\hat{\bm{k}})\equiv \delta_{ij}-\hat{k}_i\hat{k}_j$ is a projection tensor reflecting the divergence-free nature of PMFs. The first and second terms in the bracket represent non-helical and helical contributions, respectively. For later convenience, we introduce the tensor power spectrum $\mathcal{P}_{ij}(\bm{k})$ to simplify the expressions. We model the non-helical and helical PMFs with the power-law power spectrum:
\begin{align}
P_B(k) &= A_B k^{n_B}, \\
P_{H}(k) &= r_H A_Bk^{n_{H}},
\end{align}
where $n_B$ and $n_H$ are the spectral indices of the non-helical and helical PMFs, respectively.
As for the amplitudes of the power spectra, here,
$A_B$ is the amplitude of the non-helical PMFs, and
$r_H$ is introduced to represent the ratio of the helical to the non-helical PMFs (dubbed the helical-to-non-helical ratio).
Unlike the amplitude of the non-helical PMFs, $A_B$, that of the helical PMFs, $r_H A_{B}$, takes both positive and negative values.
Note that, from the Schwarz inequality for PMFs:
\begin{align}
    \lim_{\bm{k}'\to-\bm{k}}\langle\textbf{B}(\bm{k})\cdot \textbf{B}(\bm{k}')\rangle
    \geq
    \lim_{\bm{k}'\to-\bm{k}}\lvert\langle
    \lbrack
    \hat{\bm{k}}\times\textbf{B}(\bm{k})
    \rbrack
    \cdot \textbf{B}(\bm{k}')
    \rangle\rvert,
\end{align}
the power spectra of non-helical and helical PMFs obey the following inequality:~\cite{2012JCAP...06..015S}
\begin{align}\label{eq: the relation of non-helical and helical power spectra}
    P_B(k)\geq\lvert P_{H}(k)\rvert.
\end{align}
Then, this inequality leads to the constraint on the helical-to-non-helical ratio $\left| r_H \right| \leq 1$ if the two spectral indices are equivalent as $n_B=n_{H}$. We note that the scale-invariant spectrum is realized by setting $n_{B} = n_{H}=-3$.
Conventionally, the amplitude of the non-helical PMFs is parameterized by the smoothed magnetic field strength $B_r^2$ defined as
\begin{align}
B_r^2
& = 
\int^{\infty}_{0}\frac{{\rm d}k}{2\pi^{2}} e^{-k^{2}r^{2}}P_{B}(k) \notag\\
& = \frac{A_{B}}{4\pi^{2}r^{n_{B}+3}}\Gamma\left( \frac{n_{B}+3}{2} \right),
\label{eq: def Br}
\end{align}
where $\Gamma$ is the Gamma function and $r$ is the smoothing scale.

Importantly, as the passive scalar mode (\ref{eq: passive mode curvature perturbation}) is sourced by the square of the Gaussian PMFs, the statistical properties of the passive scalar mode are highly non-Gaussian. If the non-helical and helical PMF power spectra are nearly scale-invariant, i.e., $n_{B}\approx n_{H}\approx-3$, the non-Gaussianity of the PMF anisotropic stresses closes to the so-called local type (see, e.g.,~\cite{2016PhRvD..94h3503S}). Also, it has been known that the helical PMFs can source the parity-odd signature in the cosmological observables such as the CMB anisotropies (see e.g., Refs.~\cite{Pogosian:2001np, Caprini:2003vc, Kahniashvili:2005xe, 
2012PhRvD..85h3004K, 2012JCAP...06..015S, Ballardini:2014jta, Planck:2015zrl}).
In the next section, with the expression given in Eq.~\eqref{eq: passive mode curvature perturbation}, we compute the trispectrum of the passive scalar mode sourced by the PMF anisotropic stress, and see how the parity-violating signals arising from the helical PMFs (that is, $r_H \neq 0$) are characterized in the configuration space.

\section{Trispectrum of the passive scalar mode}\label{Sec: The analytical results of the passive scalar mode trispectrum}

In this section, we derive the expressions for the trispectrum of the passive scalar mode~(\ref{eq: passive mode curvature perturbation}), taking into account both the non-helical and helical PMFs. 
Together with Eqs.~(\ref{eq: passive mode curvature perturbation}) and (\ref{power spectrum of PMFs}), we explicitly compute the trispectrum of the passive scalar mode. As seen in Eq.~(\ref{eq: pi B}), the trispectrum of $\Pi_{B\,ij}$ involves an eight-point correlation function of the PMFs $\bm{B}$, and, using Wick’s theorem for Gaussian PMFs, it can be given by a sum of 105 terms, which are proportional to the 4th power of the PMF power spectrum, that is, $\propto \mathcal{P}_{ij}^4$, in general.
However, the contributions that represent the non-Gaussianity of the passive scalar mode are captured in the so-called disconnected part, which we denote by $T_{\zeta_B}$ based on the following definition:
\begin{align}
\langle\zeta_B(\bm{k}_1)\zeta_B(\bm{k}_2)\zeta_B(\bm{k}_3)\zeta_B(\bm{k}_4)\rangle_{\rm c} &\equiv(2\pi)^3\delta_{\mathrm{D}}(\bm{k}_1+\bm{k}_2+\bm{k}_3+\bm{k}_4)T_{\zeta_B}(\bm{k}_1,\bm{k}_2,\bm{k}_3,\bm{k}_4).
\end{align}
Among the 105 terms, 48 are included in this disconnected part.

Let us briefly explain how the trispectrum of the scalar mode is sensitive to the parity-violating signal, which is the main focus in this paper. The parity transformation in Fourier space leads to the sign flip of the wave vector in the trispectrum. More explicitly, performing the parity transformation to the trispectrum yields
\begin{align}
T_{\zeta_B}(\bm{k}_{1},\bm{k}_{2},\bm{k}_{3},\bm{k}_{4})
\to T_{\zeta_B}(-\bm{k}_{1},-\bm{k}_{2},-\bm{k}_{3},-\bm{k}_{4})
= \left(T_{\zeta_B}(\bm{k}_{1},\bm{k}_{2},\bm{k}_{3},\bm{k}_{4}) \right)^{*},
\end{align}
where we used the reality condition that gives $\zeta_{B}(-\bm{k})=\zeta^{*}_{B}(\bm{k})$. If the trispectrum has only the real part, i.e., $\mathrm{Im}[T_{\zeta_B}]=0$, the trispectrum is invariant under the parity transformation, suggesting the parity-even trispectrum. On the other hand, if the non-vanishing imaginary part of the trispectrum presents, i.e., $\mathrm{Im}[T_{\zeta_B}] \neq 0$, the trispectrum changes the sign under the parity transformation, i.e., the parity-odd trispectrum~\cite{2016PhRvD..94h3503S}. Thus, the imaginary part of the trispectrum can be a unique probe of parity violation in the scalar observables such as CMB temperature and $E$-mode polarization trispectra~\cite{2016PhRvD..94h3503S, Philcox:2023ffy, Philcox:2023ypl, Philcox:2025bvj,Philcox:2025lrr,Philcox:2025wts} and galaxy number density trispectra~\cite{2021arXiv210801670P, PhysRevD.106.063501, PhysRevLett.130.201002, 2023MNRAS.522.5701H, PhysRevD.107.023523, Creque-Sarbinowski_2023}. In fact, the PMF power spectrum, $\mathcal{P}_{ij}$, includes the imaginary part characterizing the helical PMFs, which is a signature of the parity violation in the magnetogenesis models (e.g., Ref.~\cite{Garretson:1992vt} for the early work), and it is expected that non-vanishing contributions to the imaginary part of the trispectrum can be induced by the helical PMFs.

To formulate the trispectrum of the passive scalar mode,
first, let us evaluate the trispectrum of the PMF anisotropic stress $\Pi_{Bab}(\bm{k})$.
After straightforward but lengthy calculations, we found that the full expression of the trispectrum of the PMF anisotropic stress $\Braket{\Pi_{Bab}(\bm{k}_1)\Pi_{Bcd}(\bm{k}_2)\Pi_{Bef}(\bm{k}_3)\Pi_{Bgh}(\bm{k}_4)}_{\rm c}$ has three contributions:
\begin{align}
\Braket{\Pi_{Bab}(\bm{k}_1)\Pi_{Bcd}(\bm{k}_2)\Pi_{Bef}(\bm{k}_3)\Pi_{Bgh}(\bm{k}_4)}_{\rm c}
= \sum_{i={\rm A,B,C}} 
\Braket{\Pi_{Bab}(\bm{k}_1)\Pi_{Bcd}(\bm{k}_2)\Pi_{Bef}(\bm{k}_3)\Pi_{Bgh}(\bm{k}_4)}_{i},
\end{align}
where we define
\begin{align}\label{eq: PiB trispectrum A}
&\Braket{\Pi_{Bab}(\bm{k}_1)\Pi_{Bcd}(\bm{k}_2)\Pi_{Bef}(\bm{k}_3)\Pi_{Bgh}(\bm{k}_4)}_{\mathrm{A}}\notag\\
& = \frac{1}{(4\pi\rho_{\gamma,0})^4}
\left( \prod_{i=1}^{4} \int {\rm d}^3\bm{k}'_i \right)
\delta_{\mathrm{D}}(\bm{k}'_2+\bm{k}'_1-\bm{k}_1)
\delta_{\mathrm{D}}(\bm{k}'_3+\bm{k}'_4-\bm{k}_2)
\delta_{\mathrm{D}}(-\bm{k}'_2-\bm{k}'_4-\bm{k}_3)
\delta_{\mathrm{D}}(-\bm{k}'_1-\bm{k}'_3-\bm{k}_4)
\notag\\
&
\qquad \times
\frac{1}{2^4}
\Biggl[
\mathcal{P}_{ac}(\bm{k}'_{1})
\mathcal{P}_{be}(\bm{k}'_{2})
\mathcal{P}_{dg}(\bm{k}'_{3})
\mathcal{P}_{fh}(\bm{k}'_{4})
+\left( a \leftrightarrow b \,\,\, \mathrm{or} \,\,\,
c \leftrightarrow d \,\,\, \mathrm{or} \,\,\,
e \leftrightarrow f \,\,\, \mathrm{or} \,\,\,
g \leftrightarrow h \right)
\Biggr]
, \\
& \Braket{\Pi_{Bab}(\bm{k}_1)\Pi_{Bcd}(\bm{k}_2)\Pi_{Bef}(\bm{k}_3)\Pi_{Bgh}(\bm{k}_4)}_{\mathrm{B}}\notag\\
& 
=\frac{1}{(4\pi\rho_{\gamma,0})^4}
\left( \prod_{i=1}^{4} \int {\rm d}^3\bm{k}'_i \right)
\delta_{\mathrm{D}}(\bm{k}'_2+\bm{k}'_1-\bm{k}_1)
\delta_{\mathrm{D}}(\bm{k}'_3+\bm{k}'_4-\bm{k}_2)
\delta_{\mathrm{D}}(-\bm{k}'_2-\bm{k}'_4-\bm{k}_3)
\delta_{\mathrm{D}}(-\bm{k}'_1-\bm{k}'_3-\bm{k}_4)
\notag\\
&
\qquad \times
\frac{1}{2^4}
\Biggl[
\mathcal{P}_{ac}(\bm{k}'_{1})
\mathcal{P}_{bg}(\bm{k}'_{2})
\mathcal{P}_{de}(\bm{k}'_{3})
\mathcal{P}_{fh}(\bm{k}'_{4})
+\left( a \leftrightarrow b \,\,\, \mathrm{or} \,\,\,
c \leftrightarrow d \,\,\, \mathrm{or} \,\,\,
e \leftrightarrow f \,\,\, \mathrm{or} \,\,\,
g \leftrightarrow h \right)
\Biggr]
, \label{eq: PiB trispectrum B} \\
&\Braket{\Pi_{Bab}(\bm{k}_1)\Pi_{Bcd}(\bm{k}_2)\Pi_{Bef}(\bm{k}_3)\Pi_{Bgh}(\bm{k}_4)}_{\mathrm{C}}\notag\\
&=\frac{1}{(4\pi\rho_{\gamma,0})^4}
\left( \prod_{i=1}^{4} \int {\rm d}^3\bm{k}'_i \right)
\delta_{\mathrm{D}}(\bm{k}'_2+\bm{k}'_1-\bm{k}_1)
\delta_{\mathrm{D}}(\bm{k}'_3+\bm{k}'_4-\bm{k}_2)
\delta_{\mathrm{D}}(-\bm{k}'_1-\bm{k}'_4-\bm{k}_3)
\delta_{\mathrm{D}}(-\bm{k}'_2-\bm{k}'_3-\bm{k}_4)
\notag\\
&
\qquad \times
\frac{1}{2^4}\Biggl[
\mathcal{P}_{ae}(\bm{k}'_{1})
\mathcal{P}_{bg}(\bm{k}'_{2})
\mathcal{P}_{ch}(\bm{k}'_{3})
\mathcal{P}_{df}(\bm{k}'_{4})
+\left( a \leftrightarrow b \,\,\, \mathrm{or} \,\,\,
c \leftrightarrow d \,\,\, \mathrm{or} \,\,\,
e \leftrightarrow f \,\,\, \mathrm{or} \,\,\,
g \leftrightarrow h \right)
\Biggr]
.\label{eq: PiB trispectrum C} 
\end{align}
The parentheses in the above stand for $2^4$ permutations of the indices. These three terms are independent contributions that do not coincide with each other with the interchange of subscripts in the tensor power spectrum, $\mathcal{P}_{ab}$. 
With the expressions of the trispectrum of the PMF anisotropic stress, the trispectrum of the scalar part of the anisotropic stress defined in Eq.~(\ref{eq: scalar PiB}) is expressed by
\begin{align}
& 
\Braket{\Pi_{B}(\bm{k}_1)\Pi_{B}(\bm{k}_2)\Pi_{B}(\bm{k}_3)\Pi_{B}(\bm{k}_4)}_{\rm c}
\notag \\
&
= O^{(0)}_{ab}(\hat{\bm{k}}_{1})
O^{(0)}_{cd}(\hat{\bm{k}}_{2})
O^{(0)}_{ef}(\hat{\bm{k}}_{3})
O^{(0)}_{gh}(\hat{\bm{k}}_{4})
\Braket{\Pi_{Bab}(\bm{k}_1)\Pi_{Bcd}(\bm{k}_2)\Pi_{Bef}(\bm{k}_3)\Pi_{Bgh}(\bm{k}_4)}_{\rm c}
\notag \\
& \equiv 
(2\pi)^{3}\delta_{\rm D}(\bm{k}_{1}+\bm{k}_{2}+\bm{k}_{3}+\bm{k}_{4}) T_{\Pi_{B}}(\bm{k}_{1},\bm{k}_{2},\bm{k}_{3},\bm{k}_{4}) .
\end{align}
From Eqs.~\eqref{eq: PiB trispectrum A}, \eqref{eq: PiB trispectrum B} and \eqref{eq: PiB trispectrum C}, we derive the expression of the trispectrum:
\begin{align}
T_{\Pi_B}(\bm{k}_{1},\bm{k}_{2},\bm{k}_{3},\bm{k}_{4})
& = 
\frac{1}{\left(4\pi \rho_{\gamma,0}\right)^4} 
O^{(0)}_{ab}(\hat{\bm{k}}_{1})
O^{(0)}_{cd}(\hat{\bm{k}}_{2})
O^{(0)}_{ef}(\hat{\bm{k}}_{3})
O^{(0)}_{gh}(\hat{\bm{k}}_{4})
\notag \\
\times&
\int\frac{{\rm d}^{3}\bm{k}'_{1}}{(2\pi)^{3}}
\Biggl[
\mathcal{P}_{ac}(\bm{k}'_{1})
\mathcal{P}_{be}(\bm{k}_{1}- \bm{k}'_{1})
\mathcal{P}_{dg}(\bm{k}'_{1} + \bm{k}_{2})
\mathcal{P}_{fh}(- \bm{k}'_{1} - \bm{k}_{2} - \bm{k}_{4})
\notag \\
&
+
\mathcal{P}_{ac}(\bm{k}'_{1})
\mathcal{P}_{bg}(\bm{k}_{1} - \bm{k}'_{1})
\mathcal{P}_{de}(\bm{k}'_{1} + \bm{k}_{2})
\mathcal{P}_{fh}(\bm{k}'_{1} + \bm{k}_{2} +\bm{k}_{3})
\notag \\
&
+
\mathcal{P}_{ae}(\bm{k}'_{1})
\mathcal{P}_{bg}(\bm{k}_{1}-\bm{k}'_{1})
\mathcal{P}_{ch}(\bm{k}'_{1} - \bm{k}_{1} - \bm{k}_{4})
\mathcal{P}_{df}(- \bm{k}'_{1} - \bm{k}_{3})
\Biggr]
, \label{eq: T Pi B exact}
\end{align}
where the first, second, and third terms in square brackets come from the contributions in Eqs.~(\ref{eq: PiB trispectrum A}), (\ref{eq: PiB trispectrum B}), and (\ref{eq: PiB trispectrum C}), respectively.
We note that the trispectrum of the scalar part of the PMF anisotropic stress $T_{\Pi_B}$ is related to the one of the scalar passive mode $T_{\zeta_B}$ through Eq.~(\ref{eq: passive mode curvature perturbation}): $T_{\zeta_B}(\bm{k}_{1}+\bm{k}_{2}+\bm{k}_{3}+\bm{k}_{4}) = \mathcal{T}^4 T_{\Pi_B}(\bm{k}_{1}+\bm{k}_{2}+\bm{k}_{3}+\bm{k}_{4})$.

To facilitate the analysis, we employ a certain approximation that allows us to analytically compute Eq.~(\ref{eq: T Pi B exact}), explained as follows.
For the nearly-scale-invariant power spectra, that is, the case with $n_{B} \approx -3$ and $n_{H} \approx -3$, the dominant contributions to the $\bm{k}$-integrals in Eqs.~(\ref{eq: PiB trispectrum A}), (\ref{eq: PiB trispectrum B}), and (\ref{eq: PiB trispectrum C}) come from four poles $k_1',\, k_2',\, k_3',\, k_4'\sim 0$. Taking these poles into the calculations, we can analytically perform the $\bm{k}$-integrals. This approximation is referred to the pole approximation~\cite{2012JCAP...06..015S}, which substantially reduces the computational costs. Conducting the pole approximation to the trispectrum, we find
\begin{align}
T_{\Pi_B}&(\bm{k}_{1},\bm{k}_{2},\bm{k}_{3},\bm{k}_{4})
\simeq
\frac{\mathcal{A}_B}{\left(4\pi \rho_{\gamma,0}\right)^4}
O^{(0)}_{ab}(\hat{\bm{k}}_{1})
O^{(0)}_{cd}(\hat{\bm{k}}_{2})
O^{(0)}_{ef}(\hat{\bm{k}}_{3})
O^{(0)}_{gh}(\hat{\bm{k}}_{4})
\notag \\
& \times
\Biggl[
\Bigl(
\delta_{ac}
\mathcal{P}_{be}(\bm{k}_{1})
\mathcal{P}_{dg}(\bm{k}_{2})
\mathcal{P}_{fh}(\bm{k}_{1} + \bm{k}_{3})
+
\mathcal{P}_{ac}(\bm{k}_{1})
\delta_{be}
\mathcal{P}_{dg}(\bm{k}_{1} + \bm{k}_{2})
\mathcal{P}_{fh}(\bm{k}_{3})
\notag\\
&
+
\mathcal{P}_{ac}(-\bm{k}_{2})
\mathcal{P}_{be}(\bm{k}_{1}+\bm{k}_{2})
\delta_{dg}
\mathcal{P}_{fh}(-\bm{k}_{4})
+
\mathcal{P}_{ac}(\bm{k}_{1}+\bm{k_{3}})
\mathcal{P}_{be}(-\bm{k}_{3})
\mathcal{P}_{dg}(-\bm{k}_{4})
\delta_{fh}
\Bigr)
\notag \\
& 
+ 
\Bigl(
\delta_{ac}
\mathcal{P}_{bg}(\bm{k}_{1})
\mathcal{P}_{de}(\bm{k}_{2})
\mathcal{P}_{fh}(\bm{k}_{2} + \bm{k}_{3})
+
\mathcal{P}_{ac}(\bm{k}_{1})
\delta_{bg}
\mathcal{P}_{de}(\bm{k}_{1} + \bm{k}_{2})
\mathcal{P}_{fh}(-\bm{k}_{4})
\notag \\
& 
+
\mathcal{P}_{ac}(-\bm{k}_{2})
\mathcal{P}_{bg}(\bm{k}_{1}+\bm{k}_{2})
\delta_{de}
\mathcal{P}_{fh}(\bm{k}_{3})
+
\mathcal{P}_{ac}(\bm{k}_{1}+\bm{k}_{4})
\mathcal{P}_{bg}(-\bm{k}_{4})
\mathcal{P}_{de}(-\bm{k}_{3})
\delta_{fh}
\Bigr)
\notag \\
& 
+
\Bigl(
\delta_{ae}
\mathcal{P}_{bg}(\bm{k}_{1})
\mathcal{P}_{ch}(\bm{k}_{2}+\bm{k}_{3})
\mathcal{P}_{df}(-\bm{k}_{3})
+
\mathcal{P}_{ae}(\bm{k}_{1})
\delta_{bg}
\mathcal{P}_{ch}(-\bm{k}_{4})
\mathcal{P}_{df}(\bm{k}_{2}+\bm{k}_{4})
\notag \\
& 
+
\mathcal{P}_{ae}(\bm{k}_{1}+\bm{k}_{4})
\mathcal{P}_{bg}(-\bm{k}_{4})
\delta_{ch}
\mathcal{P}_{df}(\bm{k}_{2})
+
\mathcal{P}_{ae}(-\bm{k}_{3})
\mathcal{P}_{bg}(\bm{k}_{1}+\bm{k}_{3})
\mathcal{P}_{ch}(\bm{k}_{2})
\delta_{df}
\Bigr)
\Biggr],
\label{eq: pole approximated trispectrum}
\end{align}
where the contributions in the first, second and third parentheses in square brackets come from the contributions in Eqs.~(\ref{eq: PiB trispectrum A}), (\ref{eq: PiB trispectrum B}), and (\ref{eq: PiB trispectrum C}), respectively.
In the above, we introduced $\mathcal{A}_B$ which is
related to the $k$-integral of the tensor power spectrum as
\begin{align}
\int \frac{{\rm d}^3\bm{k}'}{(2\pi)^3}
\mathcal{P}_{ab}(\bm{k}')
&=
\mathcal{A}_{B} \delta_{ab}, 
\end{align}
where 
\begin{align}
\mathcal{A}_{B} &\equiv \frac{2}{3}\int\frac{k^{2}{\rm d}k'}{2\pi^{2}} P_{B}(k)  = 
\frac{A_{B}}{6\pi^{2}}k^{n_{B}+3}_{*}\Gamma\left( \frac{n_{B}+3}{2}\right) 
. \label{eq: mathcal A_B}
\end{align}
Here, $\Gamma(x)$ stands for the gamma function, and $k_*\equiv 10\, \mathrm{Mpc}^{-1}$ is a UV cutoff scale of the integral, introduced via the exponential factor $\exp{\left[ (-k/k_{*})^{2} \right]}$.
We note that although the tensor power spectrum $\mathcal{P}_{ab}(\bm{k}')$ contains both the non-helical and helical power spectra (see Eq.~(\ref{power spectrum of PMFs})), only the non-helical power spectrum contributes to the parameter $\mathcal{A}_{B}$.
From the expression obtained by the pole approximation~\eqref{eq: pole approximated trispectrum}, it can be seen that the trispectrum is roughly given in the form of $\sim \mathcal{A}_B \times \mathcal{P}_{ab}^3$. Since the tensor power spectrum $\mathcal{P}_{ab}$ includes the imaginary part that characterizes the helical PMFs as in Eq.~\eqref{power spectrum of PMFs}, the imaginary part of the trispectrum reflecting the parity-violating nature is expected to depend on the form $\mathcal{A}_B \times P_B^2 P_H$ or $\mathcal{A}_B \times P_H^3$. Then, we can easily see that, in fact, after taking the contraction between the tensor power spectrum and the scalar projection tensor the imaginary part of the trispectrum can be given by
scalar triple products composed of three wave vectors, e.g., $(\hat{\bm{k}}_1 \times \hat{\bm{k}}_3) \cdot (\widehat{\bm{k}_1 + \bm{k}_2})$.

\section{Shape of the trispectrum}\label{Sec: Numerical results of the passive scalar mode trispectrum}

\begin{figure}[t]
    \centering
    \includegraphics[width=0.99\linewidth]{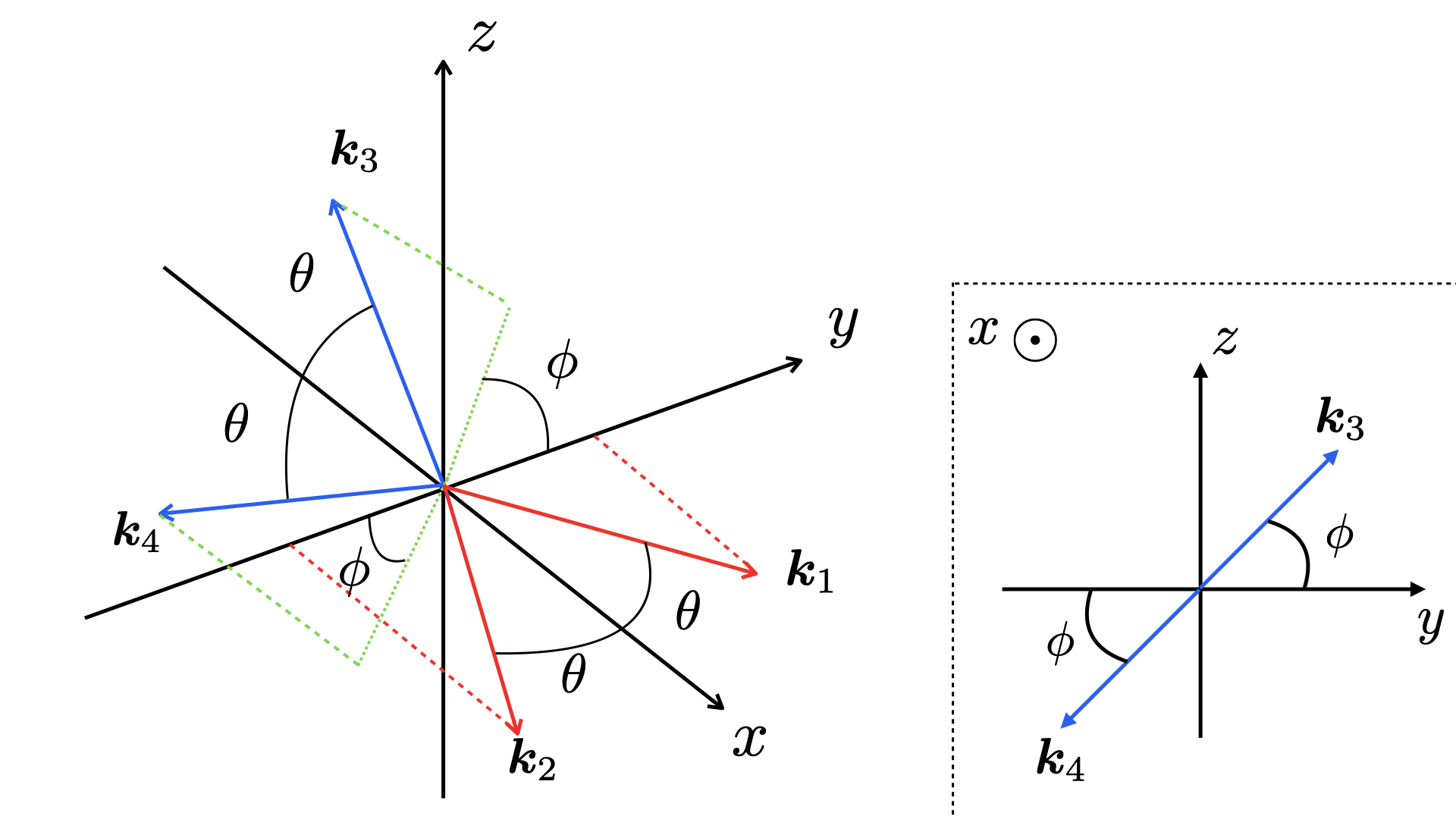}
    \caption{Equilateral configuration employed in our numerical analysis. In the right bottom panel, we show the $y$-$z$ plane, where $x$-axis is out of the page. In this configuration, the trispectrum depends on the three parameters, \{$k,\,\theta,\,\phi$\}.}
    \label{fig:equilateral conficuration}
\end{figure}

In this section, we investigate the shape of the trispectrum of the passive scalar mode $T_{\zeta_B}$ in the configuration space, by making use of Eq.~(\ref{eq: pole approximated trispectrum}).
Since the trispectrum forms a quadrangle in Fourier space, we have a high degree of freedom to specify the trispectrum.
Instead of investigating all possible quadrangles, throughout this paper we will focus on so-called the equilateral configuration~\cite{2024JCAP...05..127F}:
\begin{align}
    &\bm{k}_1=k(\cos{\theta},\, \sin{\theta},\, 0) ,\\
    &\bm{k}_2=k(\cos{\theta},\, -\sin{\theta},\, 0) ,\\
    &\bm{k}_3=k(-\cos{\theta},\, \sin{\theta}\cos{\phi},\, \sin{\theta}\sin{\phi}) ,\\
    &\bm{k}_4=k(-\cos{\theta},\, -\sin{\theta}\cos{\phi},\, -\sin{\theta}\sin{\phi}) .
\end{align}
Four wave vectors $\bm{k}_1, \bm{k}_2, \bm{k}_3$, and $\bm{k}_4$ are constructed to satisfy the momentum conservation $\bm{k}_1+\bm{k}_2+\bm{k}_3+\bm{k}_4=\bm{0}$. In this setup, all four wave vectors have the same magnitude: $k_1=k_2=k_3=k_4$. $\bm{k}_1$ and $\bm{k}_2$ lie in the $x$-$y$ plane, and $\bm{k}_3$ and $\bm{k}_4$ lie in another plane at an angle of $\phi$ to the $x$-$y$ plane around the $x$ axis. Vectors $\bm{k}_1+\bm{k}_2$ and $\bm{k}_3+\bm{k}_4$ are aligned parallel to the $x$ axis. Using this parametrization, we characterize the trispectrum by the wave number, $k$, and two angular parameters, $(\theta,\, \phi)$: $T_{\zeta_B}(\bm{k}_1,\bm{k}_2,\bm{k}_3,\bm{k}_4) = T_{\zeta_B}(k,\theta,\phi)$. Without loss of generality, we impose the conditions $0<\theta<\pi/2$ and $0\leq \phi < \pi$ by exchanging $\bm{k}_i$.
Furthermore, we introduce the dimensionless trispectrum $\tilde{T}_{\zeta_B}(k,\theta,\phi)$ defined as
\begin{align}
\tilde{T}_{\zeta_B}(k,\theta,\phi)
\equiv
k^9 \left(\frac{B^2_{r}}{8\pi\rho_{\gamma,0}}\right)^{-4} T_{\zeta_B}(k,\theta,\phi),
\end{align}
where the smoothed magnetic field strengths $B_{r}$ is defined in Eq.~(\ref{eq: def Br}). Originally, the PMF power spectra consist of four parameters, $A_{B}$, $r_{H}$, $n_{B}$ and $n_{H}$, whereas this dimensionless trispectrum depends on only three parameters, $r_{H}$, $n_{B}$ and $n_{H}$. Further setting $n_B=n_H=-2.9$, the dimensionless trispectrum depends on only the helical-to-non-helical ratio $r_{H}$. We discuss different values of $n_B$ and $n_H$ in Appendix~\ref{sec: app varying nB}.

\begin{figure}[t]
    \centering
    \includegraphics[width=\linewidth]{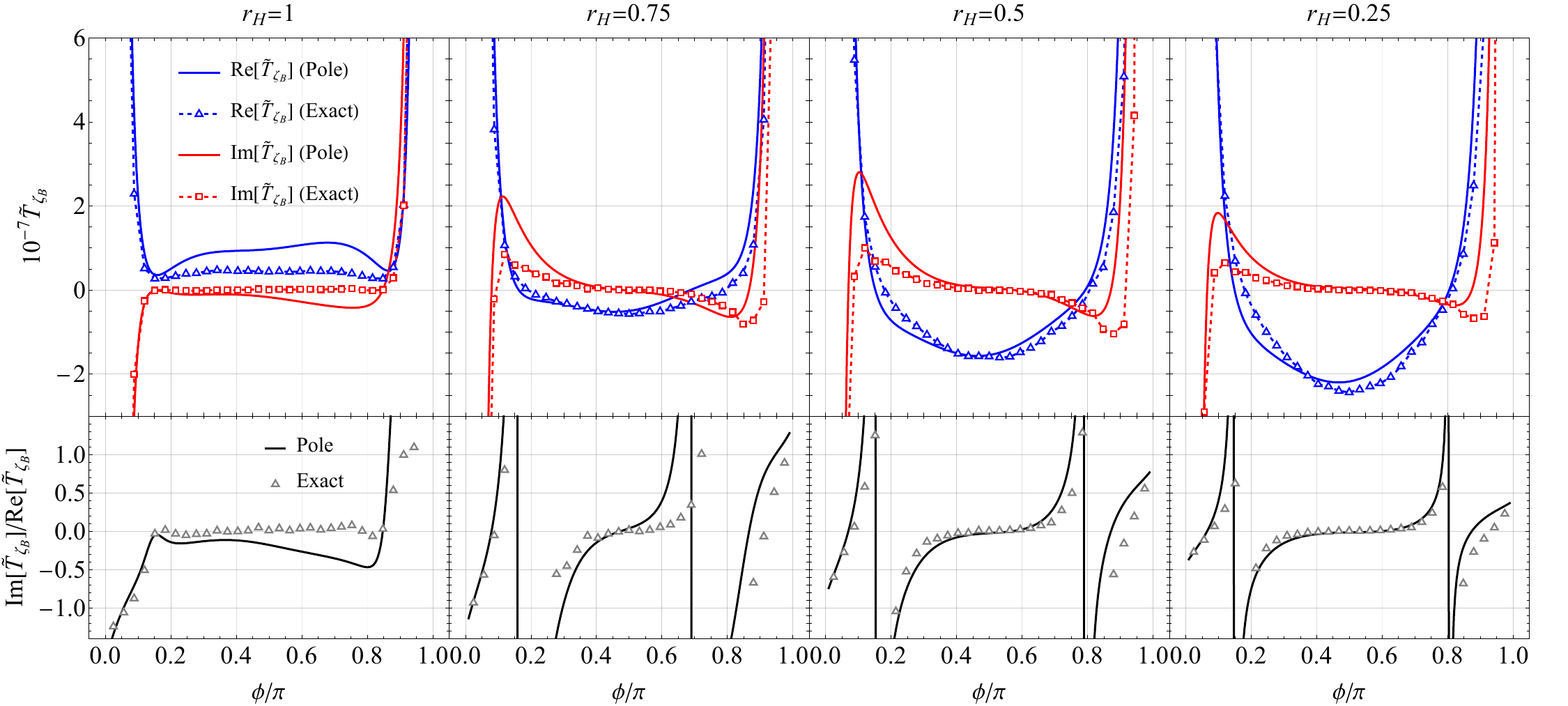}
    \caption{{\it Top}: Real (blue) and imaginary (red) components of the trispectrum at $\theta = \pi/3$, computed using the pole approximation [Eq.(\ref{eq: pole approximated trispectrum}), solid lines] and exact integration [Eq.(\ref{eq: T Pi B exact}), symbols with dashed lines] when $n_B = n_H = -2.9$. From left to right, the helical-to-non-helical ratio varies as $r_H = 1$, 0.75, 0.5, and 0.25. {\it Bottom}: Ratio of the imaginary to real components, shown for the pole approximation (solid lines) and exact integration (symbols).}
    \label{fig: pole exact phiplot}
\end{figure}

\begin{figure}[t]
    \centering
    \includegraphics[width=\linewidth]{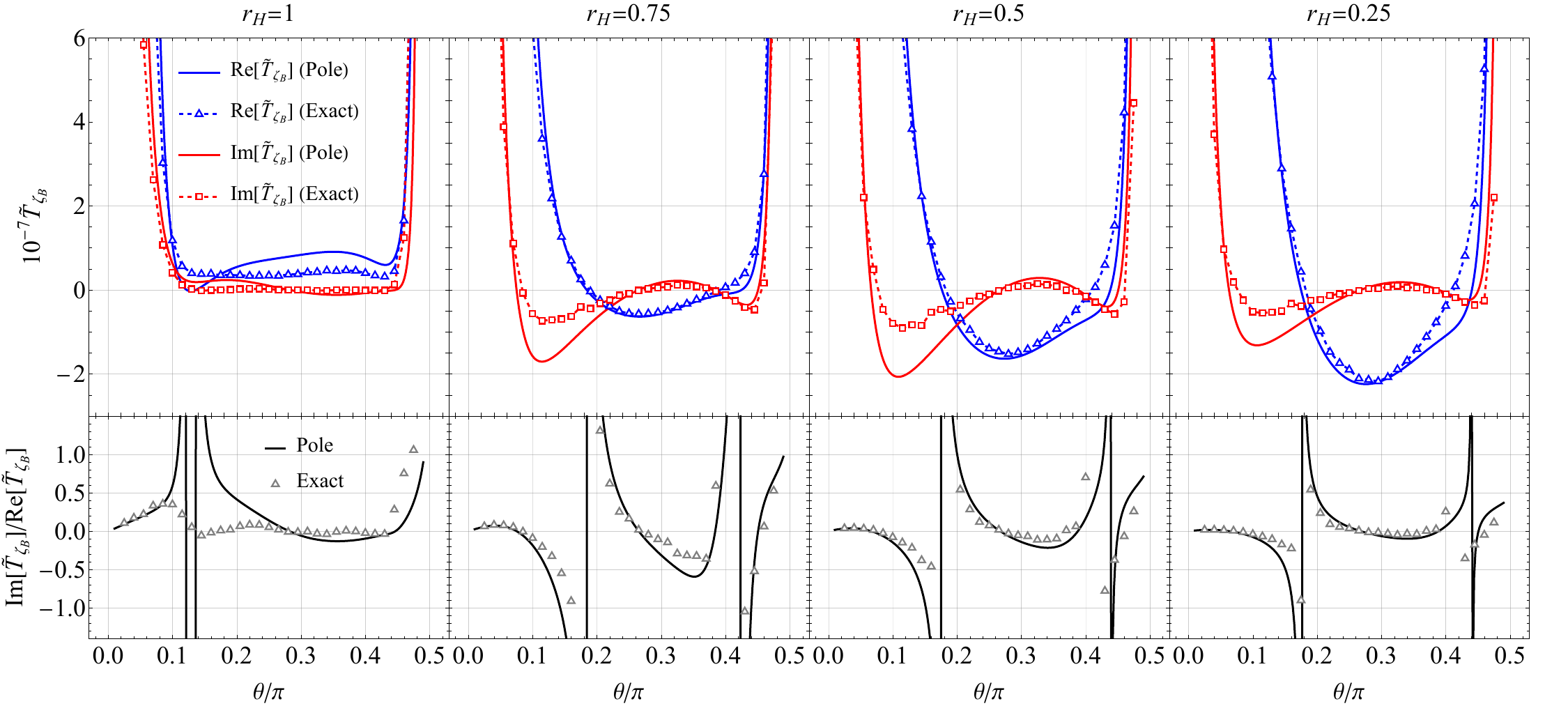}
    \caption{Same as Fig.~\ref{fig: pole exact phiplot}, but as the functions of $\theta$ with $\phi=\pi/3$.}
    \label{fig: pole exact thetaplot}
\end{figure}

We first demonstrate how the pole approximation works by comparing it with the exact integration in Figs.~\ref{fig: pole exact phiplot} and \ref{fig: pole exact thetaplot}, in which we vary $\phi$ with fixing $\theta = \pi/3$ and $\theta$ with fixing $\phi = \pi/3$, respectively. These figures show the real (parity-even) and imaginary (parity-odd) components of the trispectrum, as well as their ratio.
We first notice that a qualitative discrepancy exists for the case $r_H=1$. As the real part and imaginary part contain, respectively, the even and odd powers of $r_{H}$. Recalling Eqs.~(\ref{eq: pole approximated trispectrum}) and (\ref{eq: mathcal A_B}), the pole approximation lacks the contributions from $r^{4}_{H}$ term, leading to the deviations in the real part especially for large $r_{H}$. However, apart from the large value of $r_{H}$, the pole approximation exhibits, overall, qualitatively similar behavior to the exact results for the real and imaginary parts of the trispectrum and also the ratio of the imaginary to real components (bottom panels).

Since the trispectrum includes the PMF power spectra with the sum of two wave vectors as the argument, e.g., $\mathcal{P}_{dg} (\bm{k}_1 + \bm{k}_2)$, it has the dominant shape in the collapsed limit $\lvert \bm{k}_1+\bm{k}_2 \rvert=\lvert \bm{k}_3+\bm{k}_4 \rvert \to 0$, equivalently, $\theta \to \pi/2$. This is a well-known feature of the $\tau_{\rm NL}$-type local non-Gaussianity \cite{2016PhRvD..94h3503S} where the curvature perturbation contains chi-square fields as in our case.
Figs.~\ref{fig: pole exact phiplot} and~\ref{fig: pole exact thetaplot} show good agreement between the pole approximation and the exact integration in the collapsed limit when $n_B\approx-3$ (see also Appendix \ref{sec: app varying nB}). Our demonstration suggests that the pole approximation provides a useful trispectrum template for practically constraining the helical PMFs from CMB and galaxy observations.

\begin{figure}[t]
    \centering
    \includegraphics[width=0.99\linewidth]{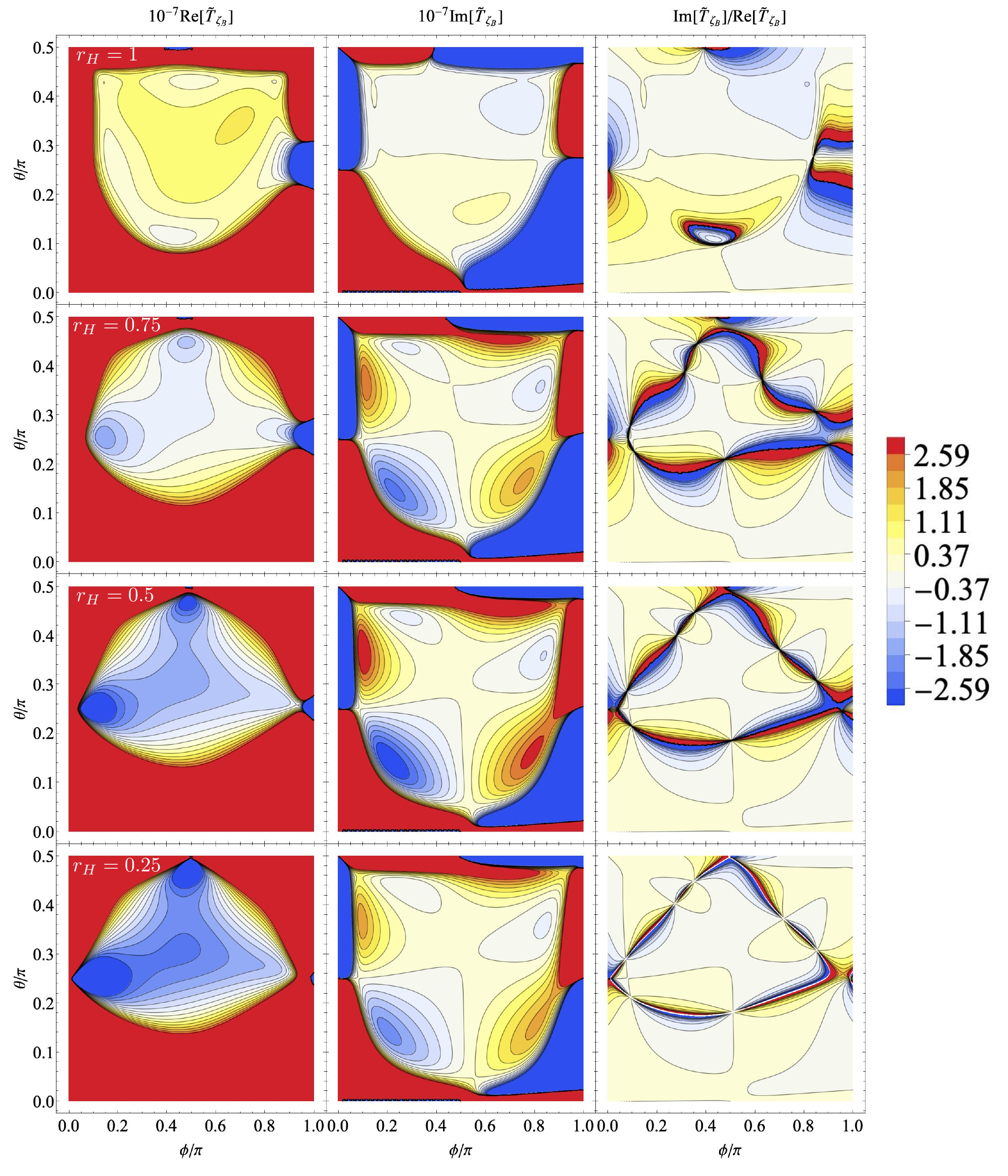}
    \caption{Trispectrum for varying the helical-to-non-helical ratio $r_{H} = 1$ to $0.25$ from top to bottom. We present the real part (left), imaginary part (center), and their ratio (right). For visualization purpose, the real and imaginary parts of the trispectrum are multiplied by $10^{-7}$. Note that the regular tetrahedron configuration ($k_{12}=k_{13}=k_{14}=k_{23}=k_{24}=k_{34}$ where $k_{ij}\equiv |\bm{k}_{i}-\bm{k}_{j}|$) corresponds to $\theta = {\rm Tan}^{-1}(\sqrt{2})\approx 0.3\pi$ and $\phi = \pi/2$.}
    \label{fig: contours}
\end{figure}

As the pole approximation significantly reduces computational cost, it enables a comprehensive exploration of the shape of the trispectrum. Focusing on the equilateral configuration, Figure~\ref{fig: contours} shows the trispectrum based on the pole approximation in the $\phi$–$\theta$ plane, varying the helical-to-non-helical ratio $r_{H}$ from 1 to 0.25 (from top to bottom panels).
Both the real and imaginary parts of the trispectrum exhibit divergent behavior for $\theta \to 0$ or $\pi/2$ and $\phi \to 0$ or $\pi$, where the configuration of wave vectors in Fourier space becomes planar. 
However, we also confirm that the imaginary part of the trispectrum vanishes exactly when the four wave vectors form a planar configuration. This is because, as discussed previously, the imaginary part is described by the terms proportional to the scalar triple product, such as $(\hat{\bm{k}}_1 \times \hat{\bm{k}}_2) \cdot (\widehat{\bm{k}_1 + \bm{k}_3})$, which becomes zero in the planar limit, due to the nature of the Levi-Civita tensor in the helical spectrum. This conclusion is consistent with the general expectation that parity-violating signals vanish when the configuration is planar.  We leave the investigation of more generic shape of the trispectrum for future full-shape analysis.
We observe that as $r_{H}$ decreases from 1 to 0.75, the structure in the $\phi$--$\theta$ plane changes qualitatively. However, for $r_{H} < 0.75$, only the amplitude changes while the overall structure remains similar. As discussed, under the pole approximation, the real part of the trispectrum contains terms proportional to $r_H^0$ and $r_H^2$, while the imaginary part contains terms proportional to $r_H^1$ and $r_H^3$. At small $r_H$, higher-order terms become subdominant, and the trispectrum is approximately proportional to $r_H^0$ (real part) and $r_H^1$ (imaginary part), explaining why its qualitative structure remains unchanged for smaller $r_H$, as seen in Fig.~\ref{fig: contours}. 
Focusing on the ratio of the imaginary to real parts (right panels), a characteristic triangular structure appears in the $\phi$–$\theta$ plane for $r_H < 0.75$.
This characteristic feature is different from the parity-violating scalar trispectrum produced by the axion dynamics during inflation, which has recently been studied in Ref.~\cite{2024JCAP...05..127F}, and
it may serve as a distinctive signature for determining the amplitude of helical PMFs from the observations.
Note that, by taking the ratio, the divergent behavior around the boundaries can be reduced, as can be seen in Fig.~\ref{fig: contours}.

\section{Implication from observations}
\label{Sec: observational constraint}

In the previous section, we have shown that the pole approximation describes the exact trispectrum especially in the collapsed limit very well.
Since the dominant signal comes from the collapsed configurations, we can exploit the pole approximation for the practical analysis. In this section, as a more practical demonstration using the pole approximation, we provide a rough constraint on the helical-to-non-helical ratio $r_{H}$ from the CMB trispectrum, referring to a template for the parity-odd trispectrum enhanced in the collapsed limit~\cite{2016PhRvD..94h3503S}: 
\begin{align}
\left[ T_{\zeta}(\bm{k}_{1},\bm{k}_{2},\bm{k}_{3},\bm{k}_{4})\right]_{d_{1}^{\rm odd}}
& \supset
-i \left\{\hat{\bm{k}}_{12}\cdot(\hat{\bm{k}}_{1}\times\hat{\bm{k}}_{3}) \right\}
\Biggl[
-  (\hat{\bm{k}}_{1}\cdot\hat{\bm{k}}_{12})
+  (\hat{\bm{k}}_{3}\cdot\hat{\bm{k}}_{12})
+  (\hat{\bm{k}}_{1}\cdot\hat{\bm{k}}_{3})
\Biggr] 
\notag \\
&
\times d_1^{\rm odd} P_{\zeta}(k_{1}) P_{\zeta}(k_{3}) P_{\zeta}(k_{12}), \label{eq: tauNL temp}
\end{align}
with $P_{\zeta}(k)$ the (scale-invariant) primordial power spectrum of the adiabatic curvature perturbations.

Picking up the corresponding term in the approximate form of the PMF-induced trispectrum explicitly shown in Appendix~\ref{sec: app pole full expresions},
we find 

\begin{align}
\left| d_{1}^{\rm odd} \right|
& = 
\mathcal{T}^{4}\frac{\mathcal{A}_B}{\left(4\pi \rho_{\gamma,0}\right)^4}
\frac{1}{81} r_{H} 
\frac{P_{B}(k_{1}) P_{B}(k_{3}) P_{B}(k_{12})}{P_{\zeta}(k_{1}) P_{\zeta}(k_{3}) P_{\zeta}(k_{12})}\notag
\\
& \approx
10\,
r_{H} 
\left(\frac{B_{r}}{\rm nG} \right)^{8} .
\end{align}
Here, for the purpose of obtaining a rough estimate, we set the scale-invariant spectrum for PMFs, i.e., $n_B = n_H = -3$, for the factors associated with $k_{1}$, $k_{3}$, and $k_{12}$, while for the terms involving Gamma functions, we simply adopt $n_{B} = n_H = -2.9$ to avoid divergences.  
Comparing this with the latest {\it Planck} limit $|d_{1}^{\rm odd}| \lesssim 10^{3}$~\cite{Philcox:2025wts},%
 \footnote{
This is obtained from the limit on $\tau_{\rm NL}^{1, \rm odd}$ in Ref.~\cite{Philcox:2025wts} via a relation $d_1^{\rm odd} = \tau_{\rm NL}^{1, \rm odd} / 6$.}
 we find
 $r_{H} \lesssim 10^{2}\left( B_{r}/{\rm nG}\right)^{-8}$. For example, taking $B_{r} = 4.7\,{\rm nG}$~\cite{Planck:2015zrl}, we have $r_{H} \lesssim 4\times 10^{-4}$. CMB measurements beyond {\it Planck} might further improve this limit by an order of magnitude \cite{2016PhRvD..94h3503S}.

 The above evaluation demonstrates a potential to constrain the helical-to-non-helical ratio through the trispectrum measurements. However, the trispectrum template we adopted \eqref{eq: tauNL temp} is too simple to recover our trispectrum entirely. For more precise discussions, a full shape analysis is necessary, and it will be done in our future work.

\section{Conclusion}
\label{Sec: conclusion}

In this work, we have presented a theoretical investigation of parity-violating signatures in the four-point correlation function in Fourier space, i.e., the trispectrum, of curvature perturbations sourced by primordial magnetic fields (PMFs). Focusing on the so-called passive scalar mode generated by the anisotropic stress of PMFs, we derived the full expression for the trispectrum, incorporating both non-helical and helical contributions. The helical component, associated with parity-violating processes in the early universe, was of particular interest, as it offers a unique window into fundamental physics beyond the standard cosmological model as well as the origin of the seed magnetic fields in the universe.

A key to investigate the parity violation in the scalar observables is that the parity-violating signals are generally encoded in the imaginary part of the trispectrum. This originates from the properties of the trispectrum under parity transformation: the real and imaginary parts are parity-even and parity-odd signatures, respectively. Since the helical component of PMFs introduces antisymmetric correlations in the magnetic field, specifically through the Levi-Civita tensor in the power spectrum (\ref{power spectrum of PMFs}), it contributes to the imaginary part of the trispectrum. This makes the imaginary component clean and unambiguous observable for detecting cosmological parity violation in the scalar sector.

To make the trispectrum analytically tractable, we employed the pole approximation, which is well-suited for nearly scale-invariant PMF spectra \cite{2012JCAP...06..015S}. Since the trispectrum, as a four-point function in Fourier space, possesses a large number of degrees of freedom in its momentum configuration, we focused our analysis on the equilateral configuration (see Fig.~\ref{fig:equilateral conficuration}) as a representative case, reducing the parameter space to manageable and allowing for a clear characterization of the signal. Then, comparing the approximate trispectrum form to exact numerical integrations, we found qualitatively good agreement in both the real (parity-even) and imaginary (parity-odd) components and also the ratio of the imaginary to real parts. Minor discrepancies in amplitude, especially for large helical-to-non-helical ratios $r_H$, were attributed to higher-order terms missed in the pole approximation.

By making use of the pole approximation, we explored the shape of the trispectrum in $\phi$--$\theta$ plane for the equilateral configuration, and examined the dependence of the real and imaginary parts on the helicity fraction $r_H$. Our analysis revealed that the imaginary component of the trispectrum shows the divergent behavior when the configuration approaches the planar structure but vanishes exactly in the planar limit, consistent with the expectation based on the symmetry-based consideration. Moreover, we found that the ratio of parity-odd to parity-even components can exceed unity for moderate values of $r_H$, providing a potentially interesting signal in forthcoming CMB and galaxy clustering surveys. Notably, the angular structure of the trispectrum exhibited distinct patterns depending on the helicity fraction $r_{H}$. For small $r_H$, the imaginary-to-real ratio displayed a characteristic triangular feature in the $\phi$--$\theta$ plane, which may serve as a distinctive template for future observational searches. As $r_H$ increases, the overall amplitude grows and the patterns evolve, arising from the nonlinear dependence of the trispectrum on the PMF helicity.

The trispectrum of the passive scalar mode has the dominant signal in the collapsed limit. 
In the above analysis, we confirmed that the pole approximation excellently recovers the collapsed-limit signal, particularly when the PMF power spectra are nearly scale-invariant. 
This allows us to apply it to more observational analyses. 
We have estimated the helical-to-non-helical ratio $r_H$ using the constraint on the trispectrum template in Ref.~\cite{2016PhRvD..94h3503S}. Assuming nearly scale-invariant PMF power spectra and the strength of the non-helical PMF of $B_{r} =4.7 \, {\rm nG}$, we obtained $r_H\lesssim 4\times 10^{-4}$.

Our results show the value of the trispectrum as a sensitive probe of helical PMFs.
The analytic and tractable pole approximation allows for efficient exploration of parameter space and will facilitate comparison with future observational data. Next generation surveys with improved measurements of CMB (e.g., LiteBIRD~\cite{LiteBIRD:2024twk}, CMB-S4~\cite{Abazajian:2019eic} and Simons observatory~\cite{SimonsObservatory:2018koc}) and galaxy clustering (e.g., SPHEREx~\cite{SPHEREx:2014bgr}, EUCLID~\cite{EUCLID:2011zbd} and PFS~\cite{PFSTeam:2012fqu}) and higher-order statistics will provide an opportunity to constrain parity-violating signatures predicted by inflationary magnetogenesis scenarios. Our framework sets the stage for such tests and motivates further theoretical and observational efforts in this direction.

\section*{Acknowledgements}
We would like to thank Tomoaki Murata and Ippei Obata for useful discussions.
This work is supported by JSPS KAKENHI Grants Nos. JP23K19050 and JP24K17043 (S.S.), Nos. JP20K03968, JP23H00108, and JP24K00627 (S.Y.), and Nos. JP20H05859 and JP23K03390 (M.S.). MS also acknowledges the Center for Computational Astrophysics, National Astronomical Observatory of Japan, for providing the computing resources.

\appendix

\section{Full expressions of the trispcetrum in the pole approximation}
\label{sec: app pole full expresions}

In this appendix, we present the explicit expressions of the trispectrum in the pole approximation, which is derived in Eq.~(\ref{eq: pole approximated trispectrum}). After properly exchanging the indices in Eq.~(\ref{eq: pole approximated trispectrum}), we have
\begin{align}
& T_{\Pi_B}(\bm{k}_{1},\bm{k}_{2},\bm{k}_{3},\bm{k}_{4})
\notag \\
& \simeq
\frac{\mathcal{A}_B}{\left(4\pi \rho_{\gamma,0}\right)^4}
O^{(0)}_{ab}(\hat{\bm{k}}_{1})
O^{(0)}_{cd}(\hat{\bm{k}}_{2})
O^{(0)}_{ef}(\hat{\bm{k}}_{3})
O^{(0)}_{gh}(\hat{\bm{k}}_{4})
\notag \\
& \times
\Biggl[
\delta_{ae}
\mathcal{P}_{cg}(\bm{k}_{12})
\mathcal{P}_{bd}(\bm{k}_{1})
\mathcal{P}_{fh}(\bm{k}_{3})
+
\delta_{cg}
\mathcal{P}_{ae}(\bm{k}_{12})
\mathcal{P}_{db}(\bm{k}_{2})
\mathcal{P}_{hf}(\bm{k}_{4})
\notag\\
&
+
\delta_{ag}
\mathcal{P}_{ce}(\bm{k}_{12})
\mathcal{P}_{bd}(\bm{k}_{1})
\mathcal{P}_{hf}(\bm{k}_{4})
+
\delta_{ce}
\mathcal{P}_{ag}(\bm{k}_{12})
\mathcal{P}_{db}(\bm{k}_{2})
\mathcal{P}_{fh}(\bm{k}_{3})
\notag \\
& 
+
\delta_{cg}
\mathcal{P}_{ea}(\bm{k}_{23})
\mathcal{P}_{df}(\bm{k}_{2})
\mathcal{P}_{hb}(\bm{k}_{4})
+
\delta_{ae}
\mathcal{P}_{cg}(\bm{k}_{23})
\mathcal{P}_{bh}(\bm{k}_{1})
\mathcal{P}_{fd}(\bm{k}_{3})
\notag\\
&
+ 
\delta_{ac}
\mathcal{P}_{eg}(\bm{k}_{23})
\mathcal{P}_{bh}(\bm{k}_{1})
\mathcal{P}_{df}(\bm{k}_{2})
+
\delta_{eg}
\mathcal{P}_{ca}(\bm{k}_{23})
\mathcal{P}_{fd}(\bm{k}_{3})
\mathcal{P}_{hb}(\bm{k}_{4})
\notag\\
&
+
\delta_{ac}
\mathcal{P}_{eg}(\bm{k}_{13})
\mathcal{P}_{bf}(\bm{k}_{1})
\mathcal{P}_{dh}(\bm{k}_{2})
+
\delta_{eg}
\mathcal{P}_{ac}(\bm{k}_{13})
\mathcal{P}_{fb}(\bm{k}_{3})
\mathcal{P}_{hd}(\bm{k}_{4})
\notag\\
&
+
\delta_{ag}
\mathcal{P}_{ec}(\bm{k}_{13})
\mathcal{P}_{bf}(\bm{k}_{1})
\mathcal{P}_{hd}(\bm{k}_{4})
+
\delta_{ce}
\mathcal{P}_{ag}(\bm{k}_{13})
\mathcal{P}_{dh}(\bm{k}_{2})
\mathcal{P}_{fb}(\bm{k}_{3})
\Biggr]. \label{eq: app pole trispectrum}
\end{align}
Using the following relation, 
\begin{align}
O^{(0)}_{ia}(\hat{\bm{k}})\mathcal{P}_{aj}(\bm{k})
& = 
\frac{1}{3}\mathcal{P}_{ij}(\bm{k}) .
\end{align}
we further simplify the expression (\ref{eq: app pole trispectrum}) as
\begin{align}
& T_{\Pi_B}(\bm{k}_{1},\bm{k}_{2},\bm{k}_{3},\bm{k}_{4})
=
\frac{\mathcal{A}_B}{\left(4\pi \rho_{\gamma,0}\right)^4}
\frac{1}{9}
\notag \\
& 
\times \Biggl[
\Biggl( 
O^{(0)}_{ab}(\hat{\bm{k}}_{2})
O^{(0)}_{cd}(\hat{\bm{k}}_{4})
\mathcal{P}_{eb}(\bm{k}_{1})
\mathcal{P}_{ed}(\bm{k}_{3})
+ (\bm{k}_{1}\leftrightarrow\bm{k}_{2})
+ (\bm{k}_{3}\leftrightarrow\bm{k}_{4})
+ (\bm{k}_{1}\leftrightarrow\bm{k}_{2}, \bm{k}_{3}\leftrightarrow\bm{k}_{4})
\Biggr)
\mathcal{P}_{ac}(\bm{k}_{12})
\Biggr]
\notag \\
& 
+
\Biggl[ (\bm{k}_{1}\to\bm{k}_{2}\to\bm{k}_{3}\to\bm{k}_{4}\to\bm{k}_1)\Biggr]
+ 
\Biggl[ \bm{k}_{2}\leftrightarrow\bm{k}_{3}\Biggr] , \label{eq: app trispectrum reduced}
\end{align}
where $(\bm{k}_{i} \leftrightarrow \bm{k}_{j})$ stands for the exchange of $\bm{k}_{i}$ and $\bm{k}_{j}$. Finally, we need to compute only one term in Eq.~(\ref{eq: app trispectrum reduced}), which is given by
\begin{align}
& 
O^{(0)}_{ab}(\hat{\bm{k}}_{2})
O^{(0)}_{cd}(\hat{\bm{k}}_{4})
\mathcal{P}_{eb}(\bm{k}_{1})
\mathcal{P}_{ed}(\bm{k}_{3})
\mathcal{P}_{ac}(\bm{k}_{12})
\notag \\
& =
\Biggl[
d_{BBB} P_B(k_{1}) P_B(k_{3}) P_B(k_{12})
+ d_{HHB} P_{H}(k_{1}) P_{H}(k_{3}) P_B(k_{12})
\notag \\
& 
+ d_{BHH} P_B(k_{1}) P_{H}(k_{3}) P_{H}(k_{12})
+ d_{HBH} P_{H}(k_{1}) P_B(k_{3}) P_{H}(k_{12})
\Biggr]
\notag \\
& 
+ i \Biggl[ d_{BHB} P_B(k_{1}) P_{H}(k_{3}) P_B(k_{12})
+ d_{HBB} P_{H}(k_{1}) P_B(k_{3}) P_B(k_{12})
\notag \\
& 
+ d_{BBH} P_B(k_{1}) P_B(k_{3}) P_{H}(k_{12})
+ d_{HHH} P_{H}(k_{1}) P_{H}(k_{3}) P_{H}(k_{12}) \Biggr]
,
\end{align}
where we define
\begin{align}
d_{BBB} &= O_{ab}(\hat{\bm{k}}_{2})O_{cd}(\hat{\bm{k}}_{4})
\left( \delta_{eb}-\hat{k}_{1e}\hat{k}_{1b}\right)
\left( \delta_{ed}-\hat{k}_{3e}\hat{k}_{3d}\right)
\left( \delta_{ac}-\hat{k}_{12a}\hat{k}_{12c}\right)
, \\
d_{HHB} &= 
- O_{ab}(\hat{\bm{k}}_{2})O_{cd}(\hat{\bm{k}}_{4})
(\delta_{bd}(\hat{\bm{k}}_{1}\cdot\hat{\bm{k}}_{3})-\hat{k}_{1d}\hat{k}_{3b})
\left( \delta_{ac}-\hat{k}_{12a}\hat{k}_{12c}\right)
, \\
d_{BHH} &= - 
O_{ab}(\hat{\bm{k}}_{2})O_{cd}(\hat{\bm{k}}_{4})
\eta_{edj}\hat{k}_{3j}\eta_{ack}\hat{k}_{12k}
\left( \delta_{eb}-\hat{k}_{1e}\hat{k}_{1b}\right)
,\\
d_{HBH} &= - 
O_{ab}(\hat{\bm{k}}_{2})O_{cd}(\hat{\bm{k}}_{4})
\eta_{ebi}\hat{k}_{1i}\eta_{ack}\hat{k}_{12k}
\left( \delta_{ed}-\hat{k}_{3e}\hat{k}_{3d}\right)
,\\
d_{BHB} &= 
O_{ab}(\hat{\bm{k}}_{2})O_{cd}(\hat{\bm{k}}_{4})
\eta_{edj}\hat{k}_{3j}
\left( \delta_{eb}-\hat{k}_{1e}\hat{k}_{1b}\right)
\left( \delta_{ac}-\hat{k}_{12a}\hat{k}_{12c}\right)
,\\
d_{HBB} &= 
O_{ab}(\hat{\bm{k}}_{2})O_{cd}(\hat{\bm{k}}_{4})
\eta_{ebi}\hat{k}_{1i}
\left( \delta_{ed}-\hat{k}_{3e}\hat{k}_{3d}\right)
\left( \delta_{ac}-\hat{k}_{12a}\hat{k}_{12c}\right)
,\\
d_{BBH} &= 
O_{ab}(\hat{\bm{k}}_{2})O_{cd}(\hat{\bm{k}}_{4})
\eta_{ack}\hat{k}_{12k}
\left( \delta_{eb}-\hat{k}_{1e}\hat{k}_{1b}\right)
\left( \delta_{ed}-\hat{k}_{3e}\hat{k}_{3d}\right)
,\\
d_{HHH} &= - 
O_{ab}(\hat{\bm{k}}_{2})O_{cd}(\hat{\bm{k}}_{4})
\eta_{ack}\hat{k}_{12k}
(\delta_{bd} (\hat{\bm{k}}_{1}\cdot\hat{\bm{k}}_{3})-\hat{k}_{1d}\hat{k}_{3b}) .
\end{align}
After lengthly but straightforward calculations, for the real part of the trispectrum, we find 
\begin{align}
d_{BBB} & =
\frac{1}{9} \Biggl[
- 6
+ (\hat{\bm{k}}_{1}\cdot\hat{\bm{k}}_{12})^2
+ (\hat{\bm{k}}_{3}\cdot\hat{\bm{k}}_{12})^2
+ (\hat{\bm{k}}_{1}\cdot\hat{\bm{k}}_{3})^2
+ 3 (\hat{\bm{k}}_{3}\cdot\hat{\bm{k}}_{4})^2
+ 3 (\hat{\bm{k}}_{4}\cdot\hat{\bm{k}}_{12})^2
\notag \\
&
+ 3 (\hat{\bm{k}}_{1}\cdot\hat{\bm{k}}_{4})^2
+ 3 (\hat{\bm{k}}_{2}\cdot\hat{\bm{k}}_{12})^2
+ 3 (\hat{\bm{k}}_{2}\cdot\hat{\bm{k}}_{3})^2
+ 3 (\hat{\bm{k}}_{1}\cdot\hat{\bm{k}}_{2})^2
+ 9 (\hat{\bm{k}}_{2}\cdot\hat{\bm{k}}_{4})^2
\notag \\
&
- (\hat{\bm{k}}_{1}\cdot\hat{\bm{k}}_{12}) (\hat{\bm{k}}_{1}\cdot\hat{\bm{k}}_{3}) (\hat{\bm{k}}_{3}\cdot\hat{\bm{k}}_{12})
- 3 (\hat{\bm{k}}_{1}\cdot\hat{\bm{k}}_{12}) (\hat{\bm{k}}_{1}\cdot\hat{\bm{k}}_{2}) (\hat{\bm{k}}_{2}\cdot\hat{\bm{k}}_{12})
- 3 (\hat{\bm{k}}_{1}\cdot\hat{\bm{k}}_{12}) (\hat{\bm{k}}_{1}\cdot\hat{\bm{k}}_{4}) (\hat{\bm{k}}_{4}\cdot\hat{\bm{k}}_{12})
\notag \\
&
- 3 (\hat{\bm{k}}_{1}\cdot\hat{\bm{k}}_{2}) (\hat{\bm{k}}_{1}\cdot\hat{\bm{k}}_{3}) (\hat{\bm{k}}_{2}\cdot\hat{\bm{k}}_{3})
- 3 (\hat{\bm{k}}_{1}\cdot\hat{\bm{k}}_{3}) (\hat{\bm{k}}_{1}\cdot\hat{\bm{k}}_{4}) (\hat{\bm{k}}_{3}\cdot\hat{\bm{k}}_{4})
\notag \\
& 
- 3 (\hat{\bm{k}}_{2}\cdot\hat{\bm{k}}_{12}) (\hat{\bm{k}}_{2}\cdot\hat{\bm{k}}_{3}) (\hat{\bm{k}}_{3}\cdot\hat{\bm{k}}_{12})
- 3 (\hat{\bm{k}}_{3}\cdot\hat{\bm{k}}_{12}) (\hat{\bm{k}}_{3}\cdot\hat{\bm{k}}_{4}) (\hat{\bm{k}}_{4}\cdot\hat{\bm{k}}_{12})
\notag \\
&
- 9 (\hat{\bm{k}}_{1}\cdot\hat{\bm{k}}_{2}) (\hat{\bm{k}}_{1}\cdot\hat{\bm{k}}_{4}) (\hat{\bm{k}}_{2}\cdot\hat{\bm{k}}_{4})
- 9 (\hat{\bm{k}}_{2}\cdot\hat{\bm{k}}_{12}) (\hat{\bm{k}}_{2}\cdot\hat{\bm{k}}_{4}) (\hat{\bm{k}}_{4}\cdot\hat{\bm{k}}_{12})
- 9 (\hat{\bm{k}}_{2}\cdot\hat{\bm{k}}_{3}) (\hat{\bm{k}}_{2}\cdot\hat{\bm{k}}_{4}) (\hat{\bm{k}}_{3}\cdot\hat{\bm{k}}_{4})
\notag \\
&
+ 3 (\hat{\bm{k}}_{1}\cdot\hat{\bm{k}}_{12}) (\hat{\bm{k}}_{1}\cdot\hat{\bm{k}}_{3}) (\hat{\bm{k}}_{3}\cdot\hat{\bm{k}}_{4}) (\hat{\bm{k}}_{4}\cdot\hat{\bm{k}}_{12})
+ 3 (\hat{\bm{k}}_{1}\cdot\hat{\bm{k}}_{2}) (\hat{\bm{k}}_{1}\cdot\hat{\bm{k}}_{3}) (\hat{\bm{k}}_{2}\cdot\hat{\bm{k}}_{12}) (\hat{\bm{k}}_{3}\cdot\hat{\bm{k}}_{12})
\notag \\
&
+ 9 (\hat{\bm{k}}_{2}\cdot\hat{\bm{k}}_{12}) (\hat{\bm{k}}_{2}\cdot\hat{\bm{k}}_{3}) (\hat{\bm{k}}_{3}\cdot\hat{\bm{k}}_{4}) (\hat{\bm{k}}_{4}\cdot\hat{\bm{k}}_{12})
+ 9 (\hat{\bm{k}}_{1}\cdot\hat{\bm{k}}_{2}) (\hat{\bm{k}}_{1}\cdot\hat{\bm{k}}_{3}) (\hat{\bm{k}}_{2}\cdot\hat{\bm{k}}_{4}) (\hat{\bm{k}}_{3}\cdot\hat{\bm{k}}_{4})
\notag \\
&
+ 9 (\hat{\bm{k}}_{1}\cdot\hat{\bm{k}}_{2}) (\hat{\bm{k}}_{1}\cdot\hat{\bm{k}}_{4}) (\hat{\bm{k}}_{2}\cdot\hat{\bm{k}}_{12}) (\hat{\bm{k}}_{4}\cdot\hat{\bm{k}}_{12})
- 9 (\hat{\bm{k}}_{1}\cdot\hat{\bm{k}}_{2}) (\hat{\bm{k}}_{1}\cdot\hat{\bm{k}}_{3}) (\hat{\bm{k}}_{2}\cdot\hat{\bm{k}}_{12}) (\hat{\bm{k}}_{3}\cdot\hat{\bm{k}}_{4}) (\hat{\bm{k}}_{4}\cdot\hat{\bm{k}}_{12})
\Biggr] , \\
d_{HHB} & =
\frac{1}{9}
\Biggl[
+ 5 (\hat{\bm{k}}_{1}\cdot\hat{\bm{k}}_{3})
- (\hat{\bm{k}}_{1}\cdot\hat{\bm{k}}_{12}) (\hat{\bm{k}}_{3}\cdot\hat{\bm{k}}_{12})
- 3 (\hat{\bm{k}}_{1}\cdot\hat{\bm{k}}_{2}) (\hat{\bm{k}}_{2}\cdot\hat{\bm{k}}_{3})
- 3 (\hat{\bm{k}}_{1}\cdot\hat{\bm{k}}_{4}) (\hat{\bm{k}}_{3}\cdot\hat{\bm{k}}_{4})
\notag \\
&
- 3 (\hat{\bm{k}}_{1}\cdot\hat{\bm{k}}_{3}) (\hat{\bm{k}}_{2}\cdot\hat{\bm{k}}_{12})^2
- 3 (\hat{\bm{k}}_{1}\cdot\hat{\bm{k}}_{3}) (\hat{\bm{k}}_{4}\cdot\hat{\bm{k}}_{12})^2
- 9 (\hat{\bm{k}}_{1}\cdot\hat{\bm{k}}_{3}) (\hat{\bm{k}}_{2}\cdot\hat{\bm{k}}_{4})^2
\notag \\
&
+ 3 (\hat{\bm{k}}_{1}\cdot\hat{\bm{k}}_{12}) (\hat{\bm{k}}_{2}\cdot\hat{\bm{k}}_{12}) (\hat{\bm{k}}_{2}\cdot\hat{\bm{k}}_{3})
+ 3 (\hat{\bm{k}}_{1}\cdot\hat{\bm{k}}_{4}) (\hat{\bm{k}}_{3}\cdot\hat{\bm{k}}_{12}) (\hat{\bm{k}}_{4}\cdot\hat{\bm{k}}_{12})
\notag \\
&
+ 9 (\hat{\bm{k}}_{1}\cdot\hat{\bm{k}}_{4}) (\hat{\bm{k}}_{2}\cdot\hat{\bm{k}}_{3}) (\hat{\bm{k}}_{2}\cdot\hat{\bm{k}}_{4})
+ 9 (\hat{\bm{k}}_{1}\cdot\hat{\bm{k}}_{3}) (\hat{\bm{k}}_{2}\cdot\hat{\bm{k}}_{12}) (\hat{\bm{k}}_{2}\cdot\hat{\bm{k}}_{4}) (\hat{\bm{k}}_{4}\cdot\hat{\bm{k}}_{12})
\notag \\
&
- 9 (\hat{\bm{k}}_{1}\cdot\hat{\bm{k}}_{4}) (\hat{\bm{k}}_{2}\cdot\hat{\bm{k}}_{12}) (\hat{\bm{k}}_{2}\cdot\hat{\bm{k}}_{3}) (\hat{\bm{k}}_{4}\cdot\hat{\bm{k}}_{12})
\Biggr]
, \\
d_{BHH} & =
\frac{1}{9} \Biggl[ 
+ 3 \left\{ \hat{\bm{k}}_{12}\cdot(\hat{\bm{k}}_{1}\times\hat{\bm{k}}_{4}) \right\}
 \left\{ \hat{\bm{k}}_{1}\cdot(\hat{\bm{k}}_{3}\times\hat{\bm{k}}_{4}) \right\}
+9 \left\{ \hat{\bm{k}}_{12}\cdot(\hat{\bm{k}}_{2}\times\hat{\bm{k}}_{4}) \right\}
 \left\{ \hat{\bm{k}}_{2}\cdot(\hat{\bm{k}}_{3}\times\hat{\bm{k}}_{4}) \right\}
\notag \\
&
-9 \left\{ \hat{\bm{k}}_{12}\cdot(\hat{\bm{k}}_{2}\times\hat{\bm{k}}_{4}) \right\} \left\{ \hat{\bm{k}}_{1}\cdot(\hat{\bm{k}}_{3}\times\hat{\bm{k}}_{4}) \right\} (\hat{\bm{k}}_{3}\cdot\hat{\bm{k}}_{4}) 
\notag \\
&
+ 5 (\hat{\bm{k}}_{3}\cdot\hat{\bm{k}}_{12})
- (\hat{\bm{k}}_{1}\cdot\hat{\bm{k}}_{12}) (\hat{\bm{k}}_{1}\cdot\hat{\bm{k}}_{3})
- 3 (\hat{\bm{k}}_{2}\cdot\hat{\bm{k}}_{12}) (\hat{\bm{k}}_{2}\cdot\hat{\bm{k}}_{3})
- 3 (\hat{\bm{k}}_{3}\cdot\hat{\bm{k}}_{4}) (\hat{\bm{k}}_{4}\cdot\hat{\bm{k}}_{12})
\notag \\
&
+ 3 (\hat{\bm{k}}_{1}\cdot\hat{\bm{k}}_{12}) (\hat{\bm{k}}_{3}\cdot\hat{\bm{k}}_{4}) (\hat{\bm{k}}_{2}\cdot\hat{\bm{k}}_{3})
- 3 (\hat{\bm{k}}_{3}\cdot\hat{\bm{k}}_{4})^2 (\hat{\bm{k}}_{3}\cdot\hat{\bm{k}}_{12})
\Biggr], \\
d_{HBH} & =
\frac{1}{9} \Biggl[
+ 3 \left\{ \hat{\bm{k}}_{12}\cdot(\hat{\bm{k}}_{2}\times\hat{\bm{k}}_{3}) \right\} \left\{ \hat{\bm{k}}_{1}\cdot(\hat{\bm{k}}_{2}\times\hat{\bm{k}}_{3}) \right\}
+ 9 \left\{ \hat{\bm{k}}_{12}\cdot(\hat{\bm{k}}_{2}\times\hat{\bm{k}}_{4}) \right\} \left\{ \hat{\bm{k}}_{1}\cdot(\hat{\bm{k}}_{2}\times\hat{\bm{k}}_{4}) \right\}
\notag \\
&
- 9 \left\{ \hat{\bm{k}}_{12}\cdot(\hat{\bm{k}}_{2}\times\hat{\bm{k}}_{4}) \right\} \left\{ \hat{\bm{k}}_{1}\cdot(\hat{\bm{k}}_{2}\times\hat{\bm{k}}_{3}) \right\} (\hat{\bm{k}}_{3}\cdot\hat{\bm{k}}_{4})
\notag \\
&
- 5 (\hat{\bm{k}}_{1}\cdot\hat{\bm{k}}_{12})
+ (\hat{\bm{k}}_{1}\cdot\hat{\bm{k}}_{3}) (\hat{\bm{k}}_{3}\cdot\hat{\bm{k}}_{12})
+ 3 (\hat{\bm{k}}_{1}\cdot\hat{\bm{k}}_{2}) (\hat{\bm{k}}_{2}\cdot\hat{\bm{k}}_{12})
+ 3 (\hat{\bm{k}}_{1}\cdot\hat{\bm{k}}_{4}) (\hat{\bm{k}}_{4}\cdot\hat{\bm{k}}_{12})
\notag \\
&
- 3 (\hat{\bm{k}}_{1}\cdot\hat{\bm{k}}_{4}) (\hat{\bm{k}}_{3}\cdot\hat{\bm{k}}_{12}) (\hat{\bm{k}}_{3}\cdot\hat{\bm{k}}_{4})
+ 3 (\hat{\bm{k}}_{1}\cdot\hat{\bm{k}}_{12}) (\hat{\bm{k}}_{3}\cdot\hat{\bm{k}}_{4})^2
\Biggr] ,
\end{align}
and, for the imaginary part of the trispectrum, 
\begin{align}
d_{BHB} & = 
-\frac{1}{9}\left\{\hat{\bm{k}}_{12}\cdot(\hat{\bm{k}}_{1}\times\hat{\bm{k}}_{3}) \right\} (\hat{\bm{k}}_{1}\cdot\hat{\bm{k}}_{12})
-\frac{1}{3}\left\{\hat{\bm{k}}_{12}\cdot(\hat{\bm{k}}_{2}\times\hat{\bm{k}}_{3}) \right\} (\hat{\bm{k}}_{2}\cdot\hat{\bm{k}}_{12})
\notag \\
&
+\frac{1}{3}\left\{ \hat{\bm{k}}_{1}\cdot(\hat{\bm{k}}_{2}\times\hat{\bm{k}}_{3}) \right\} (\hat{\bm{k}}_{1}\cdot\hat{\bm{k}}_{2})
-\frac{1}{3}\left\{ \hat{\bm{k}}_{1}\cdot(\hat{\bm{k}}_{3}\times\hat{\bm{k}}_{4}) \right\} (\hat{\bm{k}}_{1}\cdot\hat{\bm{k}}_{4})
- \left\{ \hat{\bm{k}}_{2}\cdot(\hat{\bm{k}}_{3}\times\hat{\bm{k}}_{4}) \right\} (\hat{\bm{k}}_{2}\cdot\hat{\bm{k}}_{4})
\notag \\
&
+ \left\{ \hat{\bm{k}}_{1}\cdot(\hat{\bm{k}}_{3}\times\hat{\bm{k}}_{4}) \right\} (\hat{\bm{k}}_{1}\cdot\hat{\bm{k}}_{2}) (\hat{\bm{k}}_{2}\cdot\hat{\bm{k}}_{4})
+ \left\{ \hat{\bm{k}}_{2}\cdot(\hat{\bm{k}}_{3}\times\hat{\bm{k}}_{4}) \right\} (\hat{\bm{k}}_{2}\cdot\hat{\bm{k}}_{12}) (\hat{\bm{k}}_{4}\cdot\hat{\bm{k}}_{12})
\notag \\
&
+\frac{1}{3}\left\{ \hat{\bm{k}}_{1}\cdot(\hat{\bm{k}}_{3}\times\hat{\bm{k}}_{4}) \right\} (\hat{\bm{k}}_{1}\cdot\hat{\bm{k}}_{12}) (\hat{\bm{k}}_{4}\cdot\hat{\bm{k}}_{12})
+\frac{1}{3} \left\{ \hat{\bm{k}}_{12}\cdot(\hat{\bm{k}}_{1}\times\hat{\bm{k}}_{3}) \right\} (\hat{\bm{k}}_{1}\cdot\hat{\bm{k}}_{2}) (\hat{\bm{k}}_{2}\cdot\hat{\bm{k}}_{12})
\notag \\
&
-\left\{ \hat{\bm{k}}_{1}\cdot(\hat{\bm{k}}_{3}\times\hat{\bm{k}}_{4}) \right\} (\hat{\bm{k}}_{1}\cdot\hat{\bm{k}}_{2}) (\hat{\bm{k}}_{2}\cdot\hat{\bm{k}}_{12}) (\hat{\bm{k}}_{4}\cdot\hat{\bm{k}}_{12})
, \\
d_{HBB} & = 
+\frac{1}{9}\left\{ \hat{\bm{k}}_{12}\cdot(\hat{\bm{k}}_{1}\times\hat{\bm{k}}_{3}) \right\} (\hat{\bm{k}}_{3}\cdot\hat{\bm{k}}_{12})
+\frac{1}{3}\left\{ \hat{\bm{k}}_{12}\cdot(\hat{\bm{k}}_{1}\times\hat{\bm{k}}_{4}) \right\} (\hat{\bm{k}}_{4}\cdot\hat{\bm{k}}_{12})
\notag \\
&
-\frac{1}{3}\left\{ \hat{\bm{k}}_{1}\cdot(\hat{\bm{k}}_{2}\times\hat{\bm{k}}_{3}) \right\} (\hat{\bm{k}}_{2}\cdot\hat{\bm{k}}_{3})
+\frac{1}{3}\left\{ \hat{\bm{k}}_{1}\cdot(\hat{\bm{k}}_{3}\times\hat{\bm{k}}_{4}) \right\} (\hat{\bm{k}}_{3}\cdot\hat{\bm{k}}_{4})
-\left\{ \hat{\bm{k}}_{1}\cdot(\hat{\bm{k}}_{2}\times\hat{\bm{k}}_{4}) \right\} (\hat{\bm{k}}_{2}\cdot\hat{\bm{k}}_{4})
\notag \\
&
+\left\{ \hat{\bm{k}}_{1}\cdot(\hat{\bm{k}}_{2}\times\hat{\bm{k}}_{3}) \right\} (\hat{\bm{k}}_{2}\cdot\hat{\bm{k}}_{4}) (\hat{\bm{k}}_{3}\cdot\hat{\bm{k}}_{4})
+\left\{ \hat{\bm{k}}_{1}\cdot(\hat{\bm{k}}_{2}\times\hat{\bm{k}}_{4}) \right\} (\hat{\bm{k}}_{2}\cdot\hat{\bm{k}}_{12}) (\hat{\bm{k}}_{4}\cdot\hat{\bm{k}}_{12})
\notag \\
&
+\frac{1}{3}\left\{ \hat{\bm{k}}_{1}\cdot(\hat{\bm{k}}_{2}\times\hat{\bm{k}}_{3}) \right\} (\hat{\bm{k}}_{2}\cdot\hat{\bm{k}}_{12}) (\hat{\bm{k}}_{3}\cdot\hat{\bm{k}}_{12})
-\frac{1}{3}\left\{ \hat{\bm{k}}_{12}\cdot(\hat{\bm{k}}_{1}\times\hat{\bm{k}}_{3}) \right\} (\hat{\bm{k}}_{3}\cdot\hat{\bm{k}}_{4}) (\hat{\bm{k}}_{4}\cdot\hat{\bm{k}}_{12})
\notag \\
&
-\left\{ \hat{\bm{k}}_{1}\cdot(\hat{\bm{k}}_{2}\times\hat{\bm{k}}_{3}) \right\} (\hat{\bm{k}}_{2}\cdot\hat{\bm{k}}_{12}) (\hat{\bm{k}}_{3}\cdot\hat{\bm{k}}_{4}) (\hat{\bm{k}}_{4}\cdot\hat{\bm{k}}_{12})
, \\
d_{BBH} &=
+\frac{1}{9}\left\{ \hat{\bm{k}}_{12}\cdot(\hat{\bm{k}}_{1}\times\hat{\bm{k}}_{3}) \right\}  (\hat{\bm{k}}_{1}\cdot\hat{\bm{k}}_{3})
+\frac{1}{3}\left\{ \hat{\bm{k}}_{12}\cdot(\hat{\bm{k}}_{1}\times\hat{\bm{k}}_{4}) \right\}  (\hat{\bm{k}}_{1}\cdot\hat{\bm{k}}_{4})
\notag \\
&
+\frac{1}{3}\left\{ \hat{\bm{k}}_{12}\cdot(\hat{\bm{k}}_{2}\times\hat{\bm{k}}_{3}) \right\}  (\hat{\bm{k}}_{2}\cdot\hat{\bm{k}}_{3})
+\left\{ \hat{\bm{k}}_{12}\cdot(\hat{\bm{k}}_{2}\times\hat{\bm{k}}_{4}) \right\}  (\hat{\bm{k}}_{2}\cdot\hat{\bm{k}}_{4})
\notag \\
&
-\frac{1}{3}\left\{ \hat{\bm{k}}_{12}\cdot(\hat{\bm{k}}_{1}\times\hat{\bm{k}}_{4}) \right\}  (\hat{\bm{k}}_{1}\cdot\hat{\bm{k}}_{3}) (\hat{\bm{k}}_{3}\cdot\hat{\bm{k}}_{4})
-\frac{1}{3}\left\{ \hat{\bm{k}}_{12}\cdot(\hat{\bm{k}}_{2}\times\hat{\bm{k}}_{3}) \right\}  (\hat{\bm{k}}_{1}\cdot\hat{\bm{k}}_{2}) (\hat{\bm{k}}_{1}\cdot\hat{\bm{k}}_{3})
\notag \\
&
-\left\{ \hat{\bm{k}}_{12}\cdot(\hat{\bm{k}}_{2}\times\hat{\bm{k}}_{4}) \right\}  (\hat{\bm{k}}_{1}\cdot\hat{\bm{k}}_{2}) (\hat{\bm{k}}_{1}\cdot\hat{\bm{k}}_{4})
-\left\{ \hat{\bm{k}}_{12}\cdot(\hat{\bm{k}}_{2}\times\hat{\bm{k}}_{4}) \right\}  (\hat{\bm{k}}_{2}\cdot\hat{\bm{k}}_{3}) (\hat{\bm{k}}_{3}\cdot\hat{\bm{k}}_{4})
\notag \\
&
+\left\{ \hat{\bm{k}}_{12}\cdot(\hat{\bm{k}}_{2}\times\hat{\bm{k}}_{4}) \right\}  (\hat{\bm{k}}_{1}\cdot\hat{\bm{k}}_{2}) (\hat{\bm{k}}_{1}\cdot\hat{\bm{k}}_{3}) (\hat{\bm{k}}_{3}\cdot\hat{\bm{k}}_{4})
, \\
d_{HHH} &= 
-\frac{1}{9}\left\{ \hat{\bm{k}}_{12}\cdot(\hat{\bm{k}}_{1}\times\hat{\bm{k}}_{3}) \right\}
\notag \\
&
-\left\{ \hat{\bm{k}}_{12}\cdot(\hat{\bm{k}}_{2}\times\hat{\bm{k}}_{4}) \right\} (\hat{\bm{k}}_{1}\cdot\hat{\bm{k}}_{3}) (\hat{\bm{k}}_{2}\cdot\hat{\bm{k}}_{4})
+\left\{ \hat{\bm{k}}_{12}\cdot(\hat{\bm{k}}_{2}\times\hat{\bm{k}}_{4}) \right\} (\hat{\bm{k}}_{1}\cdot\hat{\bm{k}}_{4}) (\hat{\bm{k}}_{2}\cdot\hat{\bm{k}}_{3}) .
\end{align}

\section{Validity of the pole approximation for larger spectral indices of PMFs}
\label{sec: app varying nB}

This appendix demonstrates the validity of the pole approximation for different values of $n_{B}$ and $n_{H}$, which we fix to $-2.9$ in the main text. The pole approximation works well if the spectral indices are close to scale-invariant case, i.e., $n_B=n_H=-3$.
We show how the pole approximation gets worse as increasing $n_{B}$ and $n_{H}$.

\begin{figure}
    \centering
    \includegraphics[width=0.99\linewidth]{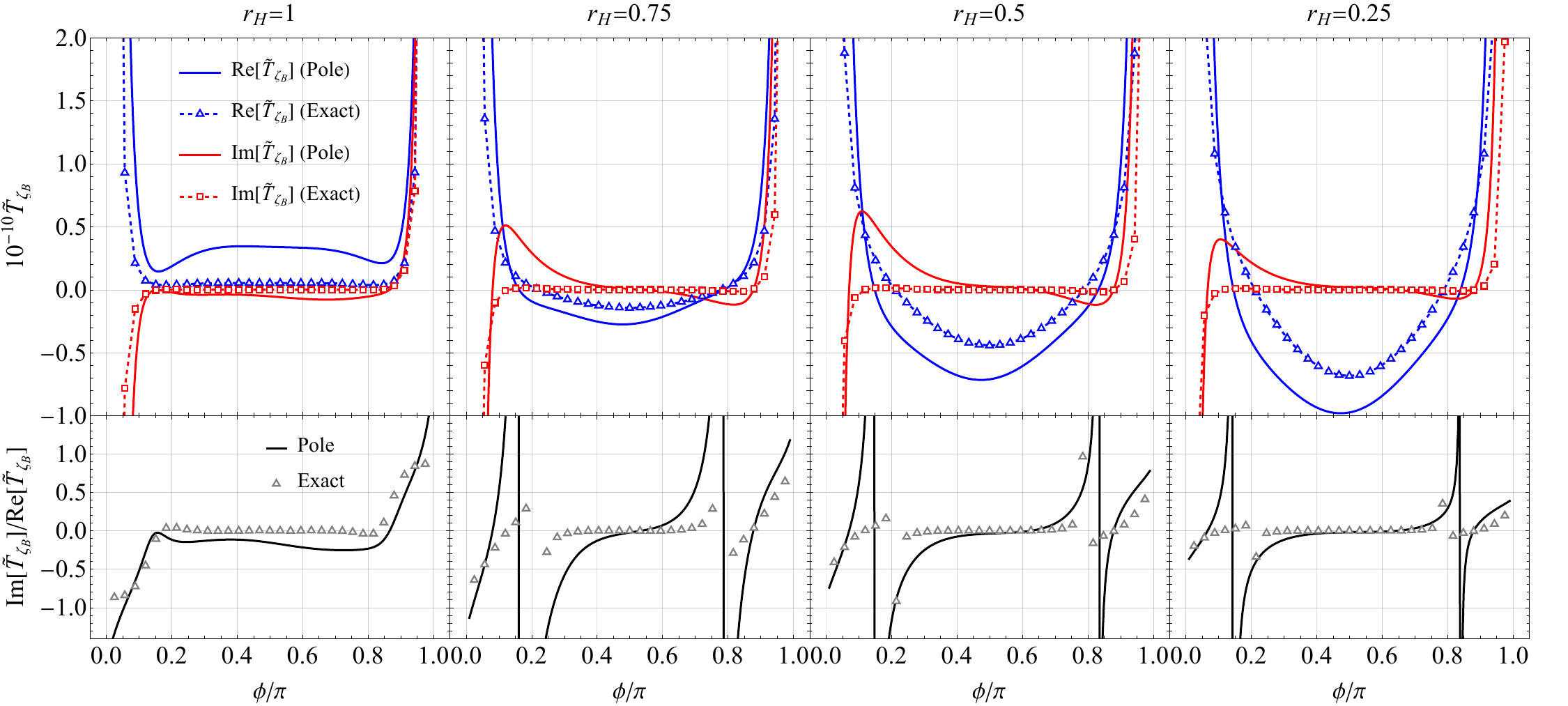}\\[1em]
    \includegraphics[width=0.99\linewidth]{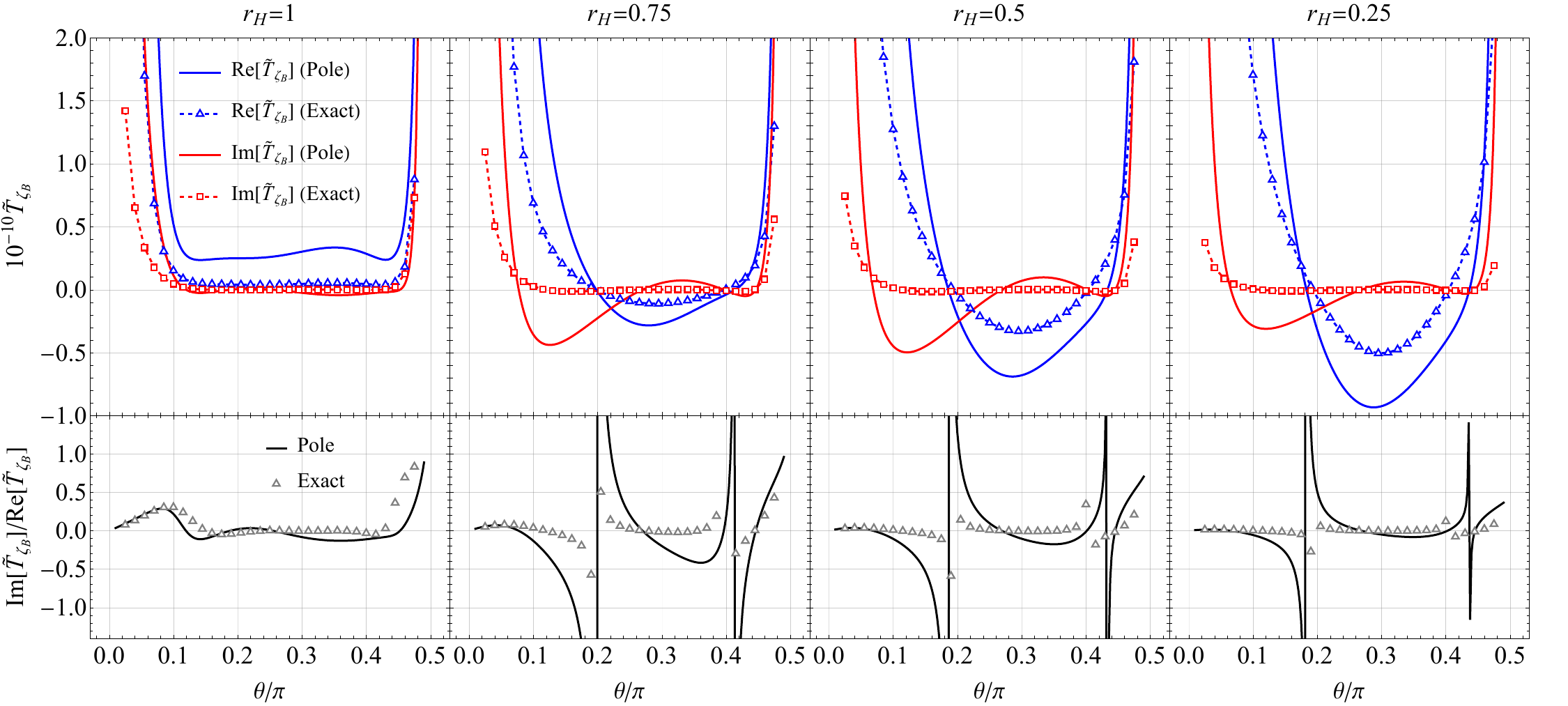}
    \caption{Same as Figs.~\ref{fig: pole exact phiplot} and \ref{fig: pole exact thetaplot} but for $n_B=n_{H}=-2.5$.}
    \label{fig: nB -2.5}
\end{figure}

\begin{figure}
    \centering
    \includegraphics[width=0.99\linewidth]{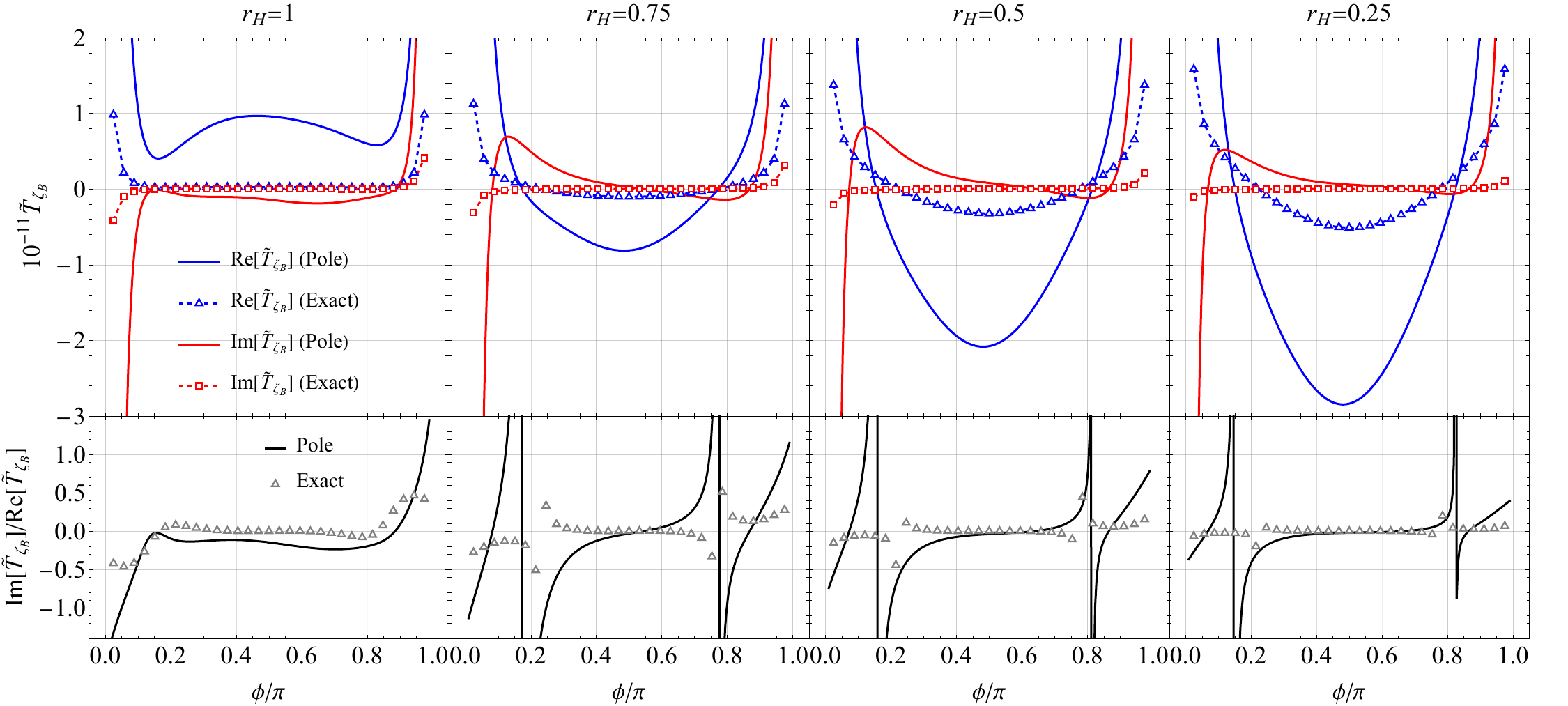}\\[1em]
    \includegraphics[width=0.99\linewidth]{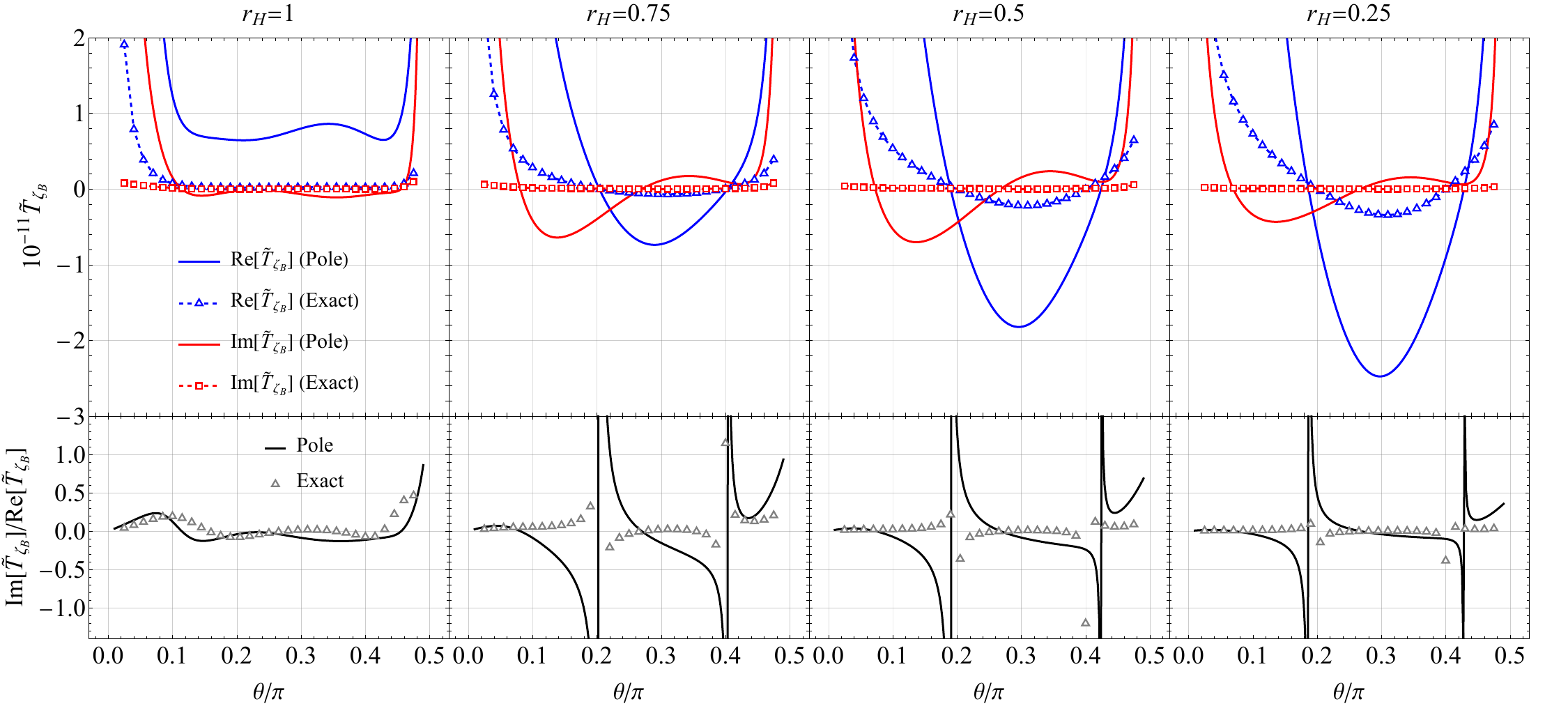}
    \caption{Same as Figs.~\ref{fig: pole exact phiplot} and \ref{fig: pole exact thetaplot} but for $n_B=n_{H}=-2$.}
    \label{fig: nB -2}
\end{figure}

Figs.~\ref{fig: nB -2.5} and \ref{fig: nB -2} show the trispectrum for $n_{B}=n_{H}=-2.5$ and $n_{B}=n_{H}=-2$, respectively. As with Figs.~\ref{fig: pole exact phiplot} and \ref{fig: pole exact thetaplot}, we present the real and imaginary parts of the trispectrum, and also their ratio, from the pole approximation and the exact integration. We clearly see that as the spectral indices increase, the disagreement between the pole approximation and the exact result becomes more prominent. Even if we look at the ratio between the real and imaginary parts of the trispectrum, the deviation is not mitigated, and hence, the validity of the pole approximation is somewhat restrictive for analyzing the trispectrum. Accordingly, our main analysis in the main text particularly focuses on the case with $n_{B}=n_{H}=-2.9$ where the pole approximation is a good approximation.

\bibliography{ref}
\bibliographystyle{JHEP}

\end{document}